\documentclass[journal,10pt]{IEEEtran}
\usepackage{amsthm}
\usepackage{amsmath,amsfonts,amssymb}
\usepackage[dvips]{graphicx}

\usepackage{subcaption}
\usepackage{verbatim}
\usepackage{setspace}
\usepackage{bm}
\usepackage{algorithmic} 
\usepackage[linesnumbered, ruled]{algorithm2e}
\SetKwRepeat{Do}{do}{while}%
\usepackage{cite}
\usepackage{caption}
\usepackage{changepage}
\usepackage{pdfpages}
\usepackage{color}
\usepackage{stfloats}
\usepackage{float}
\usepackage{graphicx}

\newtheorem{proposition}{Proposition}
\newtheorem{definition}{Definition}

\newtheorem{remark}{Remark}

\usepackage{hyperref}
\def\tr{\mathrm{tr}}
\def\re{\mathrm{Re}}

\newcommand{\rank}{\operatorname{rank}}
\usepackage{enumitem}
\newlist{diamondlist}{itemize}{1}
\setlist[diamondlist]{label=$\blacklozenge$}

\allowdisplaybreaks
\IEEEoverridecommandlockouts
\usepackage[T1]{fontenc}

\begin{document}
\title{Dual-end Fluid Antennas For Robust Anti-jamming in Low-altitude Air-ground Communications}

\author{\IEEEauthorblockN{Yifan~Guo,~Junshan~Luo,~Fanggang~Wang,~\IEEEmembership{Senior~Member,~IEEE},\\~Haiyang~Ding,~\IEEEmembership{Member,~IEEE},~Shilian~Wang,~and~Zhenhai Xu}
\vspace{-5mm}
\thanks{Yifan Guo, Junshan Luo, Shilian Wang and Zhenhai Xu are with College of Electronic Science and Technology, National University of Defense Technology, Changsha 410073, China, and Junshan Luo is also with the Sixty-Third Research Institute, National University of Defense Technology, Nanjing 210007, China (e-mails: guoyifan@nudt.edu.cn, ljsnudt@foxmail.com, wangsl@nudt.edu.cn, drxzh930@sina.com).

Fanggang Wang is with the School of Electronic and Information Engineering, Beijing Jiaotong University, Beijing 100044, China (e-mail:  wangfg@bjtu.edu.cn).

Haiyang Ding is with the School of Information and Communications, National University of Defense Technology, Wuhan 430010, China (e-mail:dinghy2003@nudt.edu.cn).
}
}
\maketitle


\begin{abstract}
This paper addresses the challenge of co-channel interference and intentional jamming in low-altitude air-ground communications.
Since conventional fixed-position antenna (FPA) systems lack spatial adaptability to dynamically balance signal enhancement against interference suppression, we propose a transformative fluid antenna system (FAS)-assisted heterogeneous dual-layer transmission architecture.
Specifically, a terrestrial base station with FPA serves ground users, while a low altitude-serving base station equipped with FAS communicates with the aerial user, also equipped with FAS, under the attack of a malicious jammer. 
We formulate a worst-case achievable rate maximization problem for aerial user subject to constraints including quality-of-service for terrestrial users, imperfect jamming directions, minimum antenna separation, etc.
To address the non-convex problem, we propose a fractional programming-block coordinate descent algorithm that alternately optimizes the transmit precoders, receive combiner, and antenna positions at both transceiver sides.
Convex hull-based approach and geometric boundary method are used to handle the jamming uncertainty and antenna placement constraints in confined spatial regions, respectively.
Extensive simulations validate significant performance gains. The FAS achieves up to 56\% higher data rates than FPA under equivalent power constraints. Strategic antenna repositioning demonstrably enhances signal quality while suppressing interference, maintaining robustness across diverse jammer channel uncertainties.
\end{abstract}

%
\IEEEpeerreviewmaketitle

\section{Introduction}
The maturation of the urban air mobility concepts and the explosive growth of applications such as drone-based logistics, surveillance, and emergency response reveal the fact that low-altitude airspace is evolving from a mere aviation resource into a pivotal economic catalyst. This emerging low-altitude economy offers the transformative potential, effectively mitigating ground traffic congestion, enhancing logistics efficiency, and revolutionizing public service delivery \cite{11017717}. Therefore, promoting the healthy development of the low-altitude economy is receiving increasing interests.

The prosperity of the low-altitude economy is profoundly dependent on establishing highly efficient and reliable air-ground communication links. Nevertheless, traditional terrestrial macro base stations, suffering from signal shadowing and cell-edge problems, exhibit inherent limitations in providing adequate aerial coverage. This pressing reality necessitates the development of novel low altitude-serving base stations (LBSs) built for the low-altitude domain, which enables precise signal beam steering and coverage optimization tailored to the unique characteristics of low-altitude user terminals \cite{ChinaTelecom2024}. Given the well-established operational ecosystem in the terrestrial networks, the prospective LBSs are likely to operate within the same frequency band. Therefore, interference from co-located terrestrial base stations (TBSs), alongside potential malicious jamming sources, drastically degrade the quality of air-ground communications.

Beamforming based on antenna array is one of the most powerful anti-jamming techniques in the spatial domain, which could nullify undesired signals from different directions \cite{8017521}. However, conventional fixed-position antenna (FPA) assumed fixed geometry and faces the trade-off between maximizing the signal power over a desired direction and minimizing the interference power over undesired directions \cite{10278220}.
Recently, researchers have proposed a novel technology of fluid antenna system (FAS) \cite{10753482} or movable antennas \cite{10906511}, which can be fabricated by surface-wave based technology \cite{9539785} or reconfigurable pixel technology \cite{6678331}.
By allowing adjusting the antenna position within a spatial region, the array geometry can be reconfigured which provides additional spatial degrees of freedom.

\subsection{Related Works}
While FAS is a well-established concept in the antenna design community, its application in wireless communications is comparatively recent. This adoption was motivated by research demonstrating that exploiting antenna position diversity offers substantial potential for performance enhancement \cite{TWC_FA_Wong}.
Current research on FAS primarily focuses on technical areas including performance analysis \cite{TWC_FA_Wong, CL_FA_Perm, TWC_FAMA, TWC_OFAMA, TWC_FA_Jake, TWC_MA_Model}, performance optimization \cite{TWC_MWY, TWC_GY, TWC_XZY}, channel estimation \cite{CL_MA_CE, CL_FA_CE, TWC_FA_CE}, physical-layer security (PLS) \cite{10912516, 10749968, 10092780, 10569014}, etc.
Specifically, the ergodic capacity and outage probability were analyzed in \cite{TWC_FA_Wong} and \cite{CL_FA_Perm}. The multiplexing gains of multi-user FASs were analyzed in \cite{TWC_FAMA} and \cite{TWC_OFAMA}. The authors of \cite{TWC_FA_Jake} proposed a generalized FAS channel model based on eigenvalue to facilitate the performance analysis of FAS. 
\cite{TWC_MA_Model} introduced a field response-based channel model, in which both the transmitter and receiver are equipped with fluid antennas, revealing the maximum achievable channel gain of FAS. 
In \cite{TWC_MWY}, the water-filling algorithm was exploited to design the transmit covariance matrix, and the successive convex approximation (SCA) algorithm was proposed to design the antenna positions of FAS. 
The SCA algorithm was further extended to multi-group multicast scenarios in \cite{TWC_GY}. The authors of \cite{TWC_XZY} introduced the particle swarm optimization algorithm for position optimization in FAS.
In addition, substantial progress has been made towards channel estimation and reconstruction for FAS, e.g., compressed sensing \cite{CL_MA_CE}, sparse channel reconstruction \cite{CL_FA_CE}, and Bayesian channel reconstruction \cite{TWC_FA_CE}.
The FAS can also be used to improve the PLS. 
For example, \cite{10912516} and \cite{10749968} deployed FAS at the transmitter, optimizing antenna positions and beamforming vector to enhance security and covertness performance. In contrast, the authors of \cite{10092780} considered using artificial random noise to disrupt eavesdropper, while the receiver employs FAS to mitigate the interference signal. \cite{10569014} further investigated the use of cooperative codewords to jam eavesdropper, enabling legitimate user to successfully decode and cancel the interference signal, while leveraging FAS to improve the secrecy rate.

In order to promote the application of FAS in low-altitude wireless networks, the following problems remain to be addressed.
First, low-altitude networks face compounded interference from terrestrial transmissions and intentional jamming. The dynamic air-ground propagation environment introduces spatial volatility that current FAS frameworks are not fully optimized to address. This scenario necessitates adaptive antenna repositioning at both LBS and aerial user to simultaneously suppress heterogeneous interferences while maintaining both terrestrial and aerial users' signal quality.
Second, given jammer's non-cooperative nature in low-altitude networks, practical deployments must accommodate jamming channel uncertainties, particularly angular ambiguities in malicious interference directions. The resulting performance degradation under bounded direction-of-arrival (DoA) errors remains unmitigated in state-of-the-art FAS designs.
Third, coordinating FAS positioning with beamforming and combining parameters presents inherent computational difficulties. The interdependence of these variables under terrestrial quality-of-service (QoS) constraints, antenna spacing limits, and jamming uncertainties creates a non-convex optimization problem, which underscores the need for novel algorithmic frameworks capable of efficient alternating optimization across interdependent system parameters.
\vspace{-3.05mm}
\subsection{Contributions}
\begin{figure}[t]\vspace{0mm}
	\begin{center}		\centerline{\includegraphics[width=0.37\textwidth]{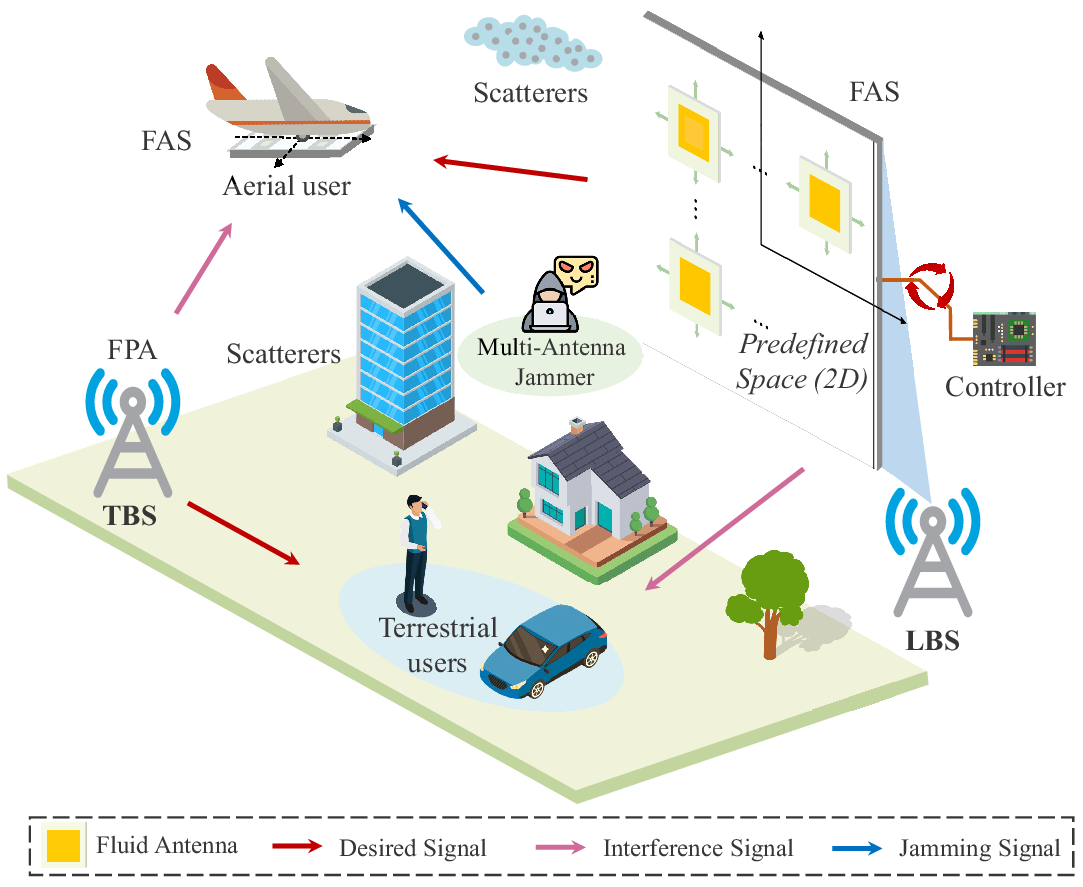}}  \vspace{-0mm}
		\captionsetup{font=footnotesize, name={Fig.}, labelsep=period}  
		\caption[t]{\raggedright FAS-assisted heterogeneous dual-layer MIMO system, where terrestrial and aerial transmission layers are equipped with distinct antenna technologies.}
		\label{fig:system}
	\end{center}
	\vspace{-6mm}
\end{figure}

Taking the above problems into account, we consider a FAS-assisted heterogeneous dual-layer MIMO system, shown as Fig. \ref{fig:system}, where a LBS serves a aerial user under the interference of a neighboring TBS and intentional jamming from a malicious jammer. 
We assume that the DoA of jamming signals can be estimated, albeit with inherent angular uncertainty. 
By jointly optimizing antenna positioning, LBS and TBS beamforming, and aerial user's combining, we aim to maximize the worst-case achievable rate of the aerial user while guaranteeing QoS for terrestrial users. 
The main contributions are as follows:
\begin{itemize}
	\item We propose a heterogeneous dual-layer communication architecture specifically designed for low-altitude networks. The system features a TBS serving ground users with conventional FPA,\footnote{The adoption of FPA for the TBS instead of FAS is primarily motivated by infrastructure legacy. The TBS operates as pre-deployed infrastructure where retrofitting FAS would incur prohibitive engineering costs and mechanical integration barriers.} complemented by a novel LBS equipped with FAS communicating with FAS-assisted aerial users. This dual-layer approach dynamically optimizes signal coverage patterns while addressing jamming threats. Compared to traditional multi-cell interference management schemes, the proposed architecture incorporates considerations of malicious jammer and angular uncertainty, which enhances robustness. More importantly, by introducing the FAS, the system shifts from passive interference suppression to active channel controlling, thereby significantly improving adaptability to the contested scenarios.
	\item To address the complex optimization challenge of coordinating antenna positions and signal processing parameters, we develop the fractional programming-block coordinate descent (FP-BCD) algorithm. This innovative approach alternates between optimizing transmitter beam directions, receiver combining schemes, and physical antenna placements. The framework employs mathematical transformations to manage QoS for terrestrial users while maximizing aerial link robustness against jamming uncertainties. A specialized geometric boundary technique with alternating optimization (AO) efficiently handles antenna positioning constraints within confined spatial regions, enabling practical implementation. For imperfect jamming DoA, the algorithm incorporates convex hull-based approach to ensure reliable performance under worst-case interference scenarios.
	\item Experimental validation demonstrates substantial performance gains over conventional systems. The FAS implementation achieves higher data rates compared to FPA systems under equivalent power constraints. Strategic antenna repositioning simultaneously enhances legitimate signal strength while suppressing interference power from jamming sources. The framework maintains excellent operational stability across different system setups. Specifically, it delivers consistent performance against malicious jamming threats, retaining performance gain even with significant jamming channel uncertainty. Radiation pattern analysis confirms precise directional nulling toward interference sources while maintaining strong signal gain toward intended aerial receivers.
\end{itemize}

\textit{Notations}: $a$, $\mathbf{a}$, $\mathbf{A}$, and $\mathcal{A}$ denote a scalar, a vector, a matrix, and a set, respectively. $(\cdot)^{\rm{T}}$, $(\cdot)^{*}$, $(\cdot)^{\rm{H}}$, $(\cdot)^{-1}$, and $(\cdot)^{\dagger}$ denote transpose, conjugate, conjugate transpose, inverse, and pseudo inverse respectively. For a matrix $\mathbf{A}$, $\mathrm{tr}(\mathbf{A})$ and $\Vert \mathbf{A} \Vert$ denotes its trace and the Frobenius norm. $\mathcal{CN}(\mathbf{0},\mathbf{\Lambda})$ denotes the circularly symmetric complex Gaussian (CSCG) distribution with mean zero and covariance matrix $\mathbf{\Lambda}$. $\mathbb{R}$ and $\mathbb{C}$ represent the sets of real and complex numbers, respectively. $\mathrm{Re}(\cdot)$, $\mathrm{Im}(\cdot)$, and $|\cdot|$ denote the real part, the imaginary part, and the amplitude of a complex number or complex vector, respectively. $\partial(\cdot)$ denotes the partial differential of a function. $\mathbf{1}_{L}$ denotes an $L$-dimensional vector with all the elements equal to 1. $\mathbf{I}_{L}$ denotes an identical matrix of size $L \times L$.

\vspace{-3mm}
\section{System Model and Problem Formulation} 
In this section, we first present an overview of the FAS-assisted heterogeneous dual-layer MIMO system. Then, we present the channel modeling for both the terrestrial and aerial links. Finally, we provide a detailed formulation of the achievable rate maximization problem.
\vspace{-5mm}
\subsection{System Model} 
As shown in Fig. \ref{fig:system}, we consider the downlink transmission in a FAS-assisted heterogeneous dual-layer MIMO system. The TBS, equipped with $N_{\mathrm{T}}$ antennas, serves $K$ single-antenna terrestrial users. On the other hand, the LBS, with $N_{\mathrm{L}}$ antennas, communicates with an aerial user that is equipped with $N_{\mathrm{A}}$ antennas. In addition, a jammer employs $N_{\mathrm{J}}$ antennas to impair the signal reception at the aerial user. 
To address the challenge of compounded interferences, both the LBS and aerial user are equipped with FAS.
This dual-ended reconfigurable architecture fundamentally transforms spatial interference management by enabling dynamic antenna repositioning at both transmission and reception sides.
We establish a 2D local coordinate system to describe the position of the $n$-th antenna for LBS, which is denoted as $\mathbf{t}_{n} = \left[x^{\mathrm{t}}_{n}, y^{\mathrm{t}}_{n}\right]^{\mathrm{T}} \in \mathcal{C}_{\mathrm{t}}$, with $n = 1, \cdots, N_{\mathrm{L}}$ and $ \mathcal{C}_{\mathrm{t}}$ being the feasible region at the transmitter.
Similarly, the position of the $m$-th antenna for aerial user is denoted as $\mathbf{r}_{m} = \left[x^{\mathrm{r}}_{m}, y^{\mathrm{r}}_{m}\right]^{\mathrm{T}} \in \mathcal{C}_{\mathrm{r}}$, with $m = 1, \cdots, N_{\mathrm{A}}$ and $\mathcal{C}_{\mathrm{r}}$ the feasible region at the receiver.
Without loss of generality, we assume that the 2D feasible moving predefined space at the LBS and aerial user are modeled as square areas with dimensions $A_{\mathrm{t}} \times A_{\mathrm{t}}$ and $A_{\mathrm{r}} \times A_{\mathrm{r}}$, respectively. 
Additionally, the position of the antenna at terrestrial user $k$ is represented by $\mathbf{u}_{k} = [0, 0]^{\mathrm{T}}, 1 \leq k \leq K$, the local coordinate of the $\tilde{n}$-th FPA at the TBS is represented by $\tilde{\mathbf{u}}_{\tilde{n}} = [X_{\tilde{n}}, Y_{\tilde{n}}]^{\mathrm{T}}, 1 \leq \tilde{n} \leq N_{\mathrm{T}}$, and the position of the $\tilde{m}$-th FPA at the jammer is represented by $\tilde{\mathbf{v}}_{\tilde{m}} = [\tilde{X}_{\tilde{m}}, \tilde{Y}_{\tilde{m}}]^{\mathrm{T}}, 1 \leq \tilde{m} \leq N_{\mathrm{J}}$.

The channels from TBS to terrestrial user $k$, from TBS to aerial user, from LBS to terrestrial user $k$, from LBS to aerial user and from jammer to aerial user are denoted as $\mathbf{h}_{\mathrm{T},k} \in \mathbb{C}^{N_{\mathrm{T}} \times 1}$, $\mathbf{H}_{\mathrm{TA}}(\tilde{\mathbf{r}}) \in \mathbb{C}^{N_{\mathrm{A}} \times N_{\mathrm{T}}}$, $\mathbf{h}_{\mathrm{L},k}(\tilde{\mathbf{r}}) \in \mathbb{C}^{N_{\mathrm{L}} \times 1}$, $\mathbf{H}_{\mathrm{LA}}(\tilde{\mathbf{t}},\tilde{\mathbf{r}}) \in \mathbb{C}^{N_{\mathrm{A}} \times N_{\mathrm{L}}}$ and $\mathbf{H}_{\mathrm{JA}}(\tilde{\mathbf{r}}) \in \mathbb{C}^{N_{\mathrm{A}} \times N_{\mathrm{J}}}$, respectively, where $\tilde{\mathbf{t}} = \left[\mathbf{t}^{\mathrm{T}}_{1}, \mathbf{t}^{\mathrm{T}}_{2}, \cdots, \mathbf{t}^{\mathrm{T}}_{N_\mathrm{L}}\right]^{\mathrm{T}}$ and $\tilde{\mathbf{r}} = \left[\mathbf{r}^{\mathrm{T}}_{1}, \mathbf{r}^{\mathrm{T}}_{2}, \cdots, \mathbf{r}^{\mathrm{T}}_{N_\mathrm{A}}\right]^{\mathrm{T}}$.
We assume that links between the terrestrial users and the jammer are obstructed by obstacles such as walls or buildings. 

The signal destined for terrestrial user $k$ is denoted as $s_{\mathrm{T},k} \!\in\! \mathbb{C}$, while the signal intended for the aerial user is denoted as $s_{\mathrm{A}} \!\in\! \mathbb{C}$. Both signals are assumed to be mutually independent, with zero mean and unit variance. 
Prior to transmission, $s_{\mathrm{T},k}$ and $s_{\mathrm{A}}$ are weighted by using transmit beamforming vectors $\mathbf{w}_{\mathrm{T},k} \in \mathbb{C}^{N_{\mathrm{T}} \times 1}$ and $\mathbf{w}_{\mathrm{L}} \in \mathbb{C}^{N_{\mathrm{L}} \times 1}$, respectively. Thus the transmit signals at the TBS and the LBS can be expressed as $\mathbf{x}_{\mathrm{T}} = \sum_{k=1}^{K}\mathbf{w}_{\mathrm{T},k}s_{\mathrm{T},k}$ and $\mathbf{x}_{\mathrm{L}} = \mathbf{w}_{\mathrm{L}} s_{\mathrm{A}}$. Meanwhile, the jammer transmits the jamming signal $\mathbf{x}_{\mathrm{J}} = \mathbf{w}_{\mathrm{J}} s_{\mathrm{J}}$ to impair the signal reception at the aerial user, where $ s_{\mathrm{J}}$ follows $\mathcal{CN}(0, 1)$ and $\mathbf{w}_{\mathrm{J}} \in \mathbb{C}^{N_{\mathrm{J}} \times 1}$ is the corresponding jamming precoder. As such, the received signal at terrestrial user $k$ is expressed as
\begin{equation}\label{eq:sig_model}
	y_{\mathrm{T},k} = \mathbf{h}^{\mathrm{H}}_{\mathrm{T},k} \mathbf{x}_{\mathrm{T}} + \mathbf{h}^{\mathrm{H}}_{\mathrm{L},k}(\tilde{\mathbf{t}})\mathbf{x}_{\mathrm{L}} + n_{\mathrm{T},k}
\end{equation}
where $n_{\mathrm{T},k} \sim \mathcal{CN}\big(0, \sigma^{2}_{\mathrm{T},k}\big)$ is the received noise of terrestrial user $k$ with variance $\sigma^{2}_{\mathrm{T},k}$.
Similarly, the received signal at the aerial user can be expressed as
\begin{equation}
y_{\mathrm{A}} \!=\! \mathbf{v}^{\mathrm{H}}_{\mathrm{A}} \big( \mathbf{H}_{\mathrm{LA}}(\tilde{\mathbf{t}}, \tilde{\mathbf{r}}) \mathbf{x}_{\mathrm{L}} \!+\! \mathbf{H}_{\mathrm{TA}}(\tilde{\mathbf{r}}) \mathbf{x}_{\mathrm{T}} \!+\! \mathbf{H}_{\mathrm{JA}}(\tilde{\mathbf{r}}) \mathbf{x}_{\mathrm{J}} \!+\! \mathbf{n}_{\mathrm{A}} \big)
\end{equation}
where $\mathbf{n}_{\mathrm{A}} \sim \mathcal{CN}\big(\bm{0}, \sigma^{2}_{\mathrm{A}}\mathbf{I}_{N_{\mathrm{A}}}\big)$ is the background noise with variance $\sigma^{2}_{\mathrm{A}}$; $\mathbf{v}_{\mathrm{A}} \in \mathbb{C}^{N_{\mathrm{A}} \times 1}$ is the combiner at the aerial user to nullify the jamming signals and balance the useful signals.
\vspace{-12mm}
\subsection{Channel Model}
In this paper, we consider field-response channel model proposed in \cite{TWC_MA_Model} and assume that the far-field condition is satisfied, where the signal propagation distance is much larger than the size of the transmit/receive region. Under this assumption, the angle-of-arrival (AoA), angle-of departure (AoD), and amplitude of the complex coefficient for each channel path remain unchanged despite the movement of the fluid antennas, which means that only the phases of multiple channel paths vary with the fluid antenna position. 

Let $L^{\mathrm{t}}_{\mathrm{L}, k}$ and $L^{\mathrm{r}}_{\mathrm{L}, k}, 1 \leq k \leq K$, denote the total number of transmit and receive channel paths from the LBS to terrestrial user $k$, respectively. Denote $\theta^{\mathrm{t}}_{\mathrm{L}, k, l} \!\in\! (-\pi/2, -\pi/2)$ and $\phi^{\mathrm{t}}_{\mathrm{L}, k, l} \!\in\! (0, \pi)$ as the elevation and azimuth AoDs of the $l$-th $(l = 1,\dots,L^{\mathrm{r}}_{\mathrm{L}, k})$ transmit path between the LBS and terrestrial user $k$, respectively. Subsequently, for terrestrial user $k$, the difference of the signal propagation distance for the $l$-th transmit channel path between $n$-th fluid antenna position $\mathbf{t}_{n}$ and the origin (i.e., $O_{\mathrm{t}} = \left[0, 0\right]^{\mathrm{T}}$) of the local coordinate system at the LBS can be represented as $\rho^{\mathrm{t}}_{\mathrm{L},k,l}\left(\mathbf{t}_{n}\right) = \mathbf{t}^{\mathrm{T}}_{n}\mathbf{n}^{\mathrm{t}}_{\mathrm{L},k,l}$, with $\mathbf{n}^{\mathrm{t}}_{\mathrm{L},k,l} = \big[\cos\theta^{\mathrm{t}}_{\mathrm{L},k,l}\cos\phi^{\mathrm{t}}_{\mathrm{L},k,l},\sin\theta^{\mathrm{t}}_{\mathrm{L},k,l}\big]^{\mathrm{T}}$ is the normalized wave vector of the $l$-th transmit path. Denoting $\lambda$ as the wavelength, the transmit field-response vector (FRV) between the LBS and terrestrial user $k$ is represented as
\begin{equation}
	\mathbf{g}_{\mathrm{L},k}(\mathbf{t}_{n}) = \big[e^{j\frac{2\pi}{\lambda}\rho^{\mathrm{t}}_{\mathrm{L},k,1}\left(\mathbf{t}_{n}\right)}, \cdots, e^{j\frac{2\pi}{\lambda}\rho^{\mathrm{t}}_{\mathrm{L},k,L^{\mathrm{t}}_{\mathrm{L}, k}}\left(\mathbf{t}_{n}\right)}\big]^{\mathrm{T}}.
\end{equation}

Then, we define the path-response matrix (PRM), $\bm{\Sigma}_{\mathrm{L}, k} \in \mathbb{C}^{L^{\mathrm{r}}_{\mathrm{L}, k} \times L^{\mathrm{t}}_{\mathrm{L}, k}}$, to represent the response between all the transmit and receive channel paths from $O_{\mathrm{t}}$ to $\mathbf{u}_{k}, 1 \leq k \leq K$. Therefore, the channel vector from the LBS to terrestrial user $k$ is obtained as
\begin{equation}\label{eq:LBS_user_channel}
	\mathbf{h}^{\mathrm{H}}_{\mathrm{L},k}(\tilde{\mathbf{t}}) = \bm{1}^{\mathrm{H}}_{L_{\mathrm{L},k}^{\mathrm{r}}}\bm{\Sigma}_{\mathrm{L},k}\mathbf{G}_{\mathrm{L},k}(\tilde{\mathbf{t}})
\end{equation}
where $\mathbf{G}_{\mathrm{L},k}(\tilde{\mathbf{t}}) \!=\! \big[\mathbf{g}_{\mathrm{L},k}(\mathbf{t}_{1}), \!\cdots\!, 	\mathbf{g}_{\mathrm{L},k}(\mathbf{t}_{N_{\mathrm{L}}})\big]$ is the field-response matrix (FRM) at the LBS. Similarly, $\mathbf{h}^{\mathrm{H}}_{\mathrm{LT},k}$ can be expressed as
\begin{equation}\label{eq:TBS_user_channel}
	\mathbf{h}^{\mathrm{H}}_{\mathrm{T},k} \!=\! \bm{1}^{\mathrm{H}}_{L_{\mathrm{T},k}^{\mathrm{r}}}\bm{\Sigma}_{\mathrm{T},k}\mathbf{G}_{\mathrm{T},k}
\end{equation}
where $\mathbf{G}_{\mathrm{T},k} \!=\! \big[\mathbf{g}_{\mathrm{T},k}(\tilde{\mathbf{u}}_{{1}}), \!\cdots\!, \mathbf{g}_{\mathrm{T},k}(\tilde{\mathbf{u}}_{N_{\mathrm{T}}})\big]$ denotes the transmit FRM at the TBS, and $\mathbf{g}_{\mathrm{T},k}(\tilde{\mathbf{u}}_{\tilde{n}}) \!=\! \big[e^{j\frac{2\pi}{\lambda}\rho^{\mathrm{t}}_{\mathrm{T},k,l}\left(\tilde{\mathbf{u}}_{\tilde{n}}\right)}\big]^{\mathrm{T}}_{1 \leq \bar{l} \leq L^{\mathrm{t}}_{\mathrm{T},k}}$. Therein, denote $ L^{\mathrm{t}}_{\mathrm{T},k}$ and $ L^{\mathrm{r}}_{\mathrm{T},k}$ as the number of transmit and receive paths from the TBS to terrestrial user $k$, respectively. In addition, $\theta^{\mathrm{t}}_{\mathrm{T}, k, \bar{l}}\in (-\pi/2, -\pi/2)$ and $\phi^{\mathrm{t}}_{\mathrm{T}, k, \bar{l}}\in (0, \pi)$ represent the elevation and azimuth AoDs for $\bar{l}$-th transmit path with $1 \leq \bar{l} \leq L^{\mathrm{t}}_{\mathrm{T},k}$, respectively. As such, we have $\rho^{\mathrm{t}}_{\mathrm{T},k,\bar{l}}\left(\tilde{\mathbf{u}}_{\tilde{n}}\right) = \tilde{\mathbf{u}}_{\tilde{n}}\mathbf{n}^{\mathrm{t}}_{\mathrm{T},k,\bar{l}}$, with $\mathbf{n}^{\mathrm{t}}_{\mathrm{T},k,\bar{l}} = \big[\cos\theta^{\mathrm{t}}_{\mathrm{T},k,\bar{l}}\cos\phi^{\mathrm{t}}_{\mathrm{T},k,\bar{l}},\sin\theta^{\mathrm{}}_{\mathrm{T},k,\bar{l}}\big]^{\mathrm{T}}$. Furthermore, $\bm{\Sigma}_{\mathrm{T},k} \in \mathbb{C}^{L^{\mathrm{r}}_{\mathrm{T},k} \times L^{\mathrm{t}}_{\mathrm{T},k}}$ represents the PRM between all transmit and receive paths.

Finally, we describe the channel matrices from the LBS, the TBS, and the jammer to the aerial user. First of all, we define $\!c \in \!\{\mathrm{LA}, \mathrm{TA}, \mathrm{JA}\}\!$ for further manipulations. Then, similar to \eqref{eq:LBS_user_channel}, the remaining channels can be expressed as
\begin{subequations}
	\begin{align}
		\mathbf{H}_{\mathrm{LA}}(\tilde{\mathbf{t}},\tilde{\mathbf{r}}) & = \mathbf{F}^{\mathrm{H}}_{\mathrm{LA}}\big(\tilde{\mathbf{r}}\big)\bm{\Sigma}_{\mathrm{LA}}\mathbf{G}_{\mathrm{LA}}(\tilde{\mathbf{t}}) \\
		\mathbf{H}_{\mathrm{TA}}(\tilde{\mathbf{r}}) & = \mathbf{F}^{\mathrm{H}}_{\mathrm{TA}}\big(\tilde{\mathbf{r}}\big)\bm{\Sigma}_{\mathrm{LA}}\mathbf{G}_{\mathrm{TA}} \\
		\mathbf{H}_{\mathrm{JA}}(\tilde{\mathbf{r}}) & = \mathbf{F}^{\mathrm{H}}_{\mathrm{JA}}\big(\tilde{\mathbf{r}}\big)\bm{\Sigma}_{\mathrm{LA}}\mathbf{G}_{\mathrm{JA}}
	\end{align}
\end{subequations}
where $\mathbf{F}_{c}(\tilde{\mathbf{r}}) \!=\! [\mathbf{f}_{c}(\mathbf{r}_{1}), \dots, \mathbf{f}_{c}(\mathbf{r}_{N_{\mathrm{A}}})]$ is the receive FRM, and $\mathbf{f}_{c}(\mathbf{r}_{m}) \!=\! \big[ e^{j \frac{2\pi}{\lambda} \rho^{\mathrm{r}}_{c, \tilde{l}}(\mathbf{r}_{m})} \big]^{\mathrm{T}}_{1 \leq \tilde{l} \leq L^{\mathrm{r}}_{c}}$, with $\!\rho^{\mathrm{r}}_{c, \tilde{l}}(\mathbf{r}_{m}) \!=\! \mathbf{r}_{m}^{\mathrm{T}} \mathbf{n}^{\mathrm{r}}_{c, \tilde{l}}$ and $\!\mathbf{n}^{\mathrm{r}}_{c, \tilde{l}} \!=\! \big[\cos \theta^{\mathrm{r}}_{c, \tilde{l}} \cos \phi^{\mathrm{r}}_{c, \tilde{l}}, \cos \theta^{\mathrm{r}}_{c, \tilde{l}}\sin \phi^{\mathrm{r}}_{c, \tilde{l}}\big]^{\mathrm{T}}$. Therein, denote $L^{\mathrm{r}}_{\mathrm{LA}}$, $L^{\mathrm{r}}_{\mathrm{TA}}$, and $L^{\mathrm{r}}_{\mathrm{JA}}$ as the number of receive paths from the LBS, TBS, and jammer to the aerial user, respectively. The elevation and azimuth AoAs are $\theta^{\mathrm{r}}_{c, \tilde{l}} \in (-\pi/2, 0)$ and $\phi^{\mathrm{r}}_{c, \tilde{l}} \in [-\pi, \pi)$ for $\tilde{l} = 1, \dots, L^{\mathrm{r}}_{c}$.
The transmit FRMs are $\mathbf{G}_{\mathrm{LA}}(\tilde{\mathbf{t}}) = [\mathbf{g}_{\mathrm{LA}}(\mathbf{t}_{1}), \dots, \mathbf{g}_{\mathrm{LA}}(\mathbf{t}_{N_{\mathrm{L}}})]$, $\mathbf{G}_{\mathrm{TA}} = [\mathbf{g}_{\mathrm{TA}}(\tilde{\mathbf{u}}_{1}), \dots, \mathbf{g}_{\mathrm{TA}}(\tilde{\mathbf{u}}_{N_{\mathrm{T}}})]$, and $\mathbf{G}_{\mathrm{JA}} = [\mathbf{g}_{\mathrm{JA}}(\tilde{\mathbf{v}}_{1}), \dots, \mathbf{g}_{\mathrm{JA}}(\tilde{\mathbf{v}}_{N_{\mathrm{J}}})]$, where
\[
\mathbf{g}_{\mathrm{LA}}(\mathbf{t}_{n}) = \big[ e^{j \frac{2\pi}{\lambda} \rho^{\mathrm{t}}_{\mathrm{LA}, l_{\mathrm{LA}}}(\mathbf{t}_{n})} \big]^{\mathrm{T}}_{1 \leq l_{\mathrm{LA}} \leq L^{\mathrm{t}}_{\mathrm{LA}}}
\]
\[
\mathbf{g}_{\mathrm{TA}}(\tilde{\mathbf{u}}_{\tilde{n}}) = \big[ e^{j \frac{2\pi}{\lambda} \rho^{\mathrm{t}}_{\mathrm{TA}, l_{\mathrm{TA}}}(\tilde{\mathbf{u}}_{\tilde{n}})} \big]^{\mathrm{T}}_{1 \leq l_{\mathrm{TA}} \leq L^{\mathrm{t}}_{\mathrm{TA}}}
\]
\[
\mathbf{g}_{\mathrm{JA}}(\tilde{\mathbf{v}}_{\tilde{m}}) = \big[ e^{j \frac{2\pi}{\lambda} \rho^{\mathrm{t}}_{\mathrm{JA}, l_{\mathrm{JA}}}(\tilde{\mathbf{v}}_{\tilde{m}})} \big]^{\mathrm{T}}_{1 \leq l_{\mathrm{JA}} \leq L^{\mathrm{t}}_{\mathrm{JA}}}
\]
with $\rho^{\mathrm{t}}_{\mathrm{LA}, l_{\mathrm{LA}}}(\mathbf{t}_{n}) = \mathbf{t}_{n}^{\mathrm{T}} \mathbf{n}^{\mathrm{t}}_{\mathrm{LA}, l_{\mathrm{LA}}}$, $\rho^{\mathrm{t}}_{\mathrm{TA}, l_{\mathrm{TA}}}(\tilde{\mathbf{u}}_{\tilde{n}}) = \tilde{\mathbf{u}}_{\tilde{n}}^{\mathrm{T}} \mathbf{n}^{\mathrm{t}}_{\mathrm{TA}, l_{\mathrm{TA}}}$, and $\rho^{\mathrm{t}}_{\mathrm{JA}, l_{\mathrm{JA}}}(\tilde{\mathbf{v}}_{\tilde{m}}) = \tilde{\mathbf{v}}_{\tilde{m}}^{\mathrm{T}} \mathbf{n}^{\mathrm{t}}_{\mathrm{JA}, l_{\mathrm{JA}}}$. Therein, denote $L^{\mathrm{t}}_{\mathrm{LA}}$, $L^{\mathrm{t}}_{\mathrm{TA}}$, and $L^{\mathrm{t}}_{\mathrm{JA}}$ as the number of transmit paths from the LBS, TBS, and jammer to the aerial user, respectively. The elevation and azimuth AoDs are $\theta^{\mathrm{t}}_{c, \hat{l}} \in (-\pi/2, \pi/2)$ and $\phi^{\mathrm{t}}_{c, \hat{l}} \in (0, \pi)$ for $\hat{l} = 1, \dots, L^{\mathrm{t}}_{c}$, with normalized wave vector:
$
\mathbf{n}^{\mathrm{t}}_{c, \hat{l}} = \big[\cos \theta^{\mathrm{t}}_{c, \hat{l}} \cos \phi^{\mathrm{t}}_{c, \hat{l}}, \sin \theta^{\mathrm{t}}_{c, \hat{l}}\big]^{\mathrm{T}}.
$
The PRMs $\bm{\Sigma}_{\mathrm{LA}} \in \mathbb{C}^{L^{\mathrm{r}}_{\mathrm{LA}} \times L^{\mathrm{t}}_{\mathrm{LA}}}$, $\bm{\Sigma}_{\mathrm{TA}} \in \mathbb{C}^{L^{\mathrm{r}}_{\mathrm{TA}} \times L^{\mathrm{t}}_{\mathrm{TA}}}$, and $\bm{\Sigma}_{\mathrm{JA}} \in \mathbb{C}^{L^{\mathrm{r}}_{\mathrm{JA}} \times L^{\mathrm{t}}_{\mathrm{JA}}}$ capture the complex gains between transmit and receive paths.
\vspace{-4mm}
\subsection{Problem Formulation}
We assume that the CSI of all the legal channels is perfectly known. The received SINR for terrestrial user $k$ is given by
\begin{equation}
	\gamma_{\mathrm{T},k} = \frac{\lvert \mathbf{h}^{\mathrm{H}}_{\mathrm{T},k}\mathbf{w}_{\mathrm{T},k} \lvert^{2}}{\sum_{i \neq k}^{K}\lvert \mathbf{h}^{\mathrm{H}}_{\mathrm{T},k}\mathbf{w}_{\mathrm{T},i} \lvert^{2} + \lvert \mathbf{h}^{\mathrm{H}}_{\mathrm{L},k}\left({\tilde{\mathbf{t}}}\right)\mathbf{w}_{\mathrm{L}} \lvert^{2} + \sigma^{2}_{\mathrm{T},k}}.
\end{equation}
The received SINR of the aerial user can be expressed as
\begin{equation}
	\begin{aligned}
		\gamma_{\mathrm{A}} \! = \! \mathbf{v}^{\mathrm{H}}_{\mathrm{A}}\mathbf{H}_{\mathrm{LA}}\left({\tilde{\mathbf{t}}},{\tilde{\mathbf{r}}}\right)\mathbf{w}_{\mathrm{L}}\mathbf{w}^{\mathrm{H}}_{\mathrm{L}}\mathbf{H}^{\mathrm{H}}_{\mathrm{LA}}\left({\tilde{\mathbf{t}}},{\tilde{\mathbf{r}}}\right)\mathbf{v}_{\mathrm{A}} \left(\mathbf{v}^{\mathrm{H}}_{\mathrm{A}}\mathbf{A}\mathbf{v}_{\mathrm{A}}\right)^{-1}
	\end{aligned}	
\end{equation}
where 
$
	\mathbf{A} \!= \!\mathbf{A}_{1}\! +\! \mathbf{H}_{\mathrm{JA}}\left({\tilde{\mathbf{r}}}\right)\mathbf{w}_{\mathrm{J}}\mathbf{w}^{\mathrm{H}}_{\mathrm{J}}\mathbf{H}^{\mathrm{H}}_{\mathrm{JA}}\left({\tilde{\mathbf{r}}}\right) \in \mathbb{C}^{N_{\mathrm{A}} \times N_{\mathrm{A}}}
$, and 
$
	\mathbf{A}_{1} \!=\! \sum_{k=1}^{K}\mathbf{H}_{\mathrm{TA}}\left({\tilde{\mathbf{r}}}\right)\mathbf{w}_{\mathrm{T},k}\mathbf{w}^{\mathrm{H}}_{\mathrm{T},k}\mathbf{H}^{\mathrm{H}}_{\mathrm{TA}}\left({\tilde{\mathbf{r}}}\right) \!+\! \sigma^2_{\mathrm{A}}\mathbf{I}_{N_{\mathrm{A}}} \in \mathbb{C}^{N_{\mathrm{A}} \times N_{\mathrm{A}}}
$.

Due to the fact that the jammer does not cooperate with the legitimate parties, the jamming CSI $\mathbf{H}_{\mathrm{JA}}$ are challenging to obtain. To account for their harmful effects on the system performance, we assume that the jammer’s CSI belongs to a given angular range ${\Delta}$ \cite{syf_ris_aj_1, syf_convex_hull}, which is 
\begin{equation}
	\Delta = \Big\{\mathbf{H}_{\mathrm{JA}} \,\big|\, \theta \in [\theta_L, \theta_U], \phi \in [\phi_L, \phi_U]\Big\}
\end{equation}
where $\theta_U$ and $\theta_L$ denote the upper and lower bounds of azimuth angle, $\phi_U$ and $\phi_L$ are the upper and lower bounds of elevation angle. Next, we formulate a worst-case achievable rate maximization problem\footnote{Since the instantaneous beamforming vector of the jammer remains unknown, we adopt the Cauchy-Schwarz inequality to rigorously upper-bound the maximum possible jamming power inflicted upon the aerial user, as in \eqref{eq:Cauchy_ineq}. This yields a tractable robust design philosophy that holds under any realizable jammer beam configuration, sacrificing marginal optimality for operational reliability in antagonistic environments.}:
\begin{subequations}
	\label{eq:opt_1}
	\begin{align}
		(\mathbf{P1}):& \ \mathop{\max}_{\tilde{\mathbf{t}}, \tilde{\mathbf{r}}, \mathbf{W}_{\mathrm{T}}, \mathbf{w}_{\mathrm{L}},  \mathbf{v}_{\mathrm{A}} } \ \mathop{\min}_{\Delta} \quad \log_{2}\left(1+\gamma_{\mathrm{A}}\right) \\
		& \qquad \text{s.t. } \tilde{\mathbf{t}} \in \mathcal{C}_{\mathrm{t}}, \tilde{\mathbf{r}} \in \mathcal{C}_{\mathrm{r}}  \label{cons:ap_size} \\
		& \qquad \phantom{\text{s.t. }}\Vert \mathbf{t}_{n_1} \!-\! \mathbf{t}_{n_2}\Vert \!\geq\! D,  1 \!\leq\! n_1 \neq n_2 \!\leq\! N_{\mathrm{T}} \label{cons:max_ap_lbs} \\
		& \qquad \phantom{\text{s.t. }}\Vert \mathbf{r}_{m_1}\! -\! \mathbf{r}_{m_2}\Vert \!\geq\! D, 1 \!\leq\! m_1 \neq m_2 \!\leq\! N_{\mathrm{A}} \label{cons:max_ap_uav} \\
		& \qquad \phantom{\text{s.t. }} \Vert \mathbf{w}_{\mathrm{L}} \Vert^{2} \leq P_{\mathrm{L,max}} \label{cons:max_power_lbs} \\
		& \qquad \phantom{\text{s.t. }} \Vert \mathbf{v}_{\mathrm{A}} \Vert^{2} = 1 \label{cons:uav_beam_unit}\\
		& \qquad \phantom{\text{s.t. }} \log_{2}\left(1+\gamma_{\mathrm{T},k}\right) \geq R_{k,\mathrm{min}}, \  \forall k \label{cons:min_user_rate}
	\end{align}
\end{subequations}
where $\mathbf{W}_{\mathrm{T}} \!=\! \left[\mathbf{w}^{\mathrm{T}}_{\mathrm{T},1}, \mathbf{w}^{\mathrm{T}}_{\mathrm{T},2}, \cdots,  \mathbf{w}^{\mathrm{T}}_{\mathrm{T},K}\right]^{\mathrm{T}} \!\in\! \mathbb{C}^{N_{\mathrm{T}} \times K}$. Constraint \eqref{cons:ap_size} indicates that the transmit and receive antennas can only move within the given regions, i.e., $\mathcal{C}_{\mathrm{t}}$ and $\mathcal{C}_{\mathrm{r}}$, respectively. Constraints \eqref{cons:max_ap_lbs} and \eqref{cons:max_ap_uav} guarantee a minimum distance $D$ between adjacent antennas, thereby avoiding coupling effects between antennas in the transmit/receive regions. Constraint \eqref{cons:max_power_lbs} is the maximum power constraint at transmitter, where $P_{\mathrm{L,max}}$ denotes the maximum transmit power, and constraint \eqref{cons:uav_beam_unit} is the receive combiner restriction, and constraint \eqref{cons:min_user_rate} is the minimum rate constraint with the $k$-th terrestrial user’s target $R_{k,\mathrm{min}}$.

\vspace{-3mm}
\section{Joint Optimization of Beamforming and Antenna Position}
In this section, we begin by transforming the optimization problem $\mathbf{P1}$ into an equivalent problem through fractional programming. Subsequently, we derive the local optimal solutions for LBS and TBS beamforming and the iterative solutions for receive combiner $\mathbf{v}_{\mathrm{A}}$, receive APV $\tilde{\mathbf{r}}$, and transmit APV $\tilde{\mathbf{t}}$. Finally, we introduce the proposed FP-BCD algorithm.
\vspace{-4mm}
\subsection{Problem Transforming}
We first apply the Lagrangian dual transformation proposed in \cite{fp_skm_tsp} to \eqref{cons:min_user_rate}, and $\mathbf{P1}$ can be equivalently written as
\begin{subequations}
	\label{eq:opt_2}
	\begin{align}
		(\mathbf{P2}):\ & \mathop{\max}_{\tilde{\mathbf{t}}, \tilde{\mathbf{r}}, \mathbf{W}_{\mathrm{T}}, \mathbf{w}_{\mathrm{L}},  \mathbf{v}_{\mathrm{A}}, \tilde{\mathbf{y}} } \ \mathop{\min}_{\Delta} \quad \log_{2}\left(1+\gamma_{\mathrm{A}}\right) \\
		& \qquad \ \ \text{s.t. } R'_{k} \geq R_{k,\mathrm{min}}, \  \forall k \label{cons:min_user_rate_2} \\
		& \qquad \ \ \phantom{\text{s.t. }} \eqref{cons:ap_size}-\eqref{cons:uav_beam_unit} \notag
	\end{align}
\end{subequations}
where $\tilde{\mathbf{y}} = \left[\tilde{y}_{1}, \tilde{y}_{2}, \cdots, \tilde{y}_{K}\right]^{\mathrm{T}} \in \mathbb{R}^{K \times 1}$ is the auxiliary variable for the SINR terms and 
$$
	R'_{k} = \log_{2}\left(1+\tilde{y}_{k}\right) - \tilde{y}_{k} + {\left(1+\tilde{y}_{k}\right)\lvert \mathbf{h}^{\mathrm{H}}_{\mathrm{T},k}\mathbf{w}_{\mathrm{T},k} \lvert^{2}}{\Gamma^{-1}_{k}}
$$
with $\Gamma_{k} =\sum_{i=1}^{K}\lvert \mathbf{h}^{\mathrm{H}}_{\mathrm{T},k}\mathbf{w}_{\mathrm{T},i} \lvert^{2} + \lvert \mathbf{h}^{\mathrm{H}}_{\mathrm{L},k}\left({\tilde{\mathbf{t}}}\right)\mathbf{w}_{\mathrm{L}} \lvert^{2} + \sigma^{2}_{\mathrm{T},k} $. Note that, for maximizing $R'_{k}$ with other variable being fixed, the optimal $\tilde{y}_{k}$ equals to the corresponding SINR term of terrestrial user $k$, i.e.,
\begin{equation}\label{eq:opt_y}
	\tilde{y}_{k} = \frac{\lvert \mathbf{h}^{\mathrm{H}}_{\mathrm{T},k}\mathbf{w}_{\mathrm{T},k} \lvert^{2}}{\sum_{i \neq k}^{K}\lvert \mathbf{h}^{\mathrm{H}}_{\mathrm{T},k}\mathbf{w}_{\mathrm{T},i} \lvert^{2} + \lvert \mathbf{h}^{\mathrm{H}}_{\mathrm{L},k}\left({\tilde{\mathbf{t}}}\right)\mathbf{w}_{\mathrm{L}} \lvert^{2} + \sigma^{2}_{\mathrm{T},k}}.
\end{equation}
After obtaining the optimal $\tilde{y}_{k}$, we can find that $R'_{k} = R_{k}$. Therefore, with the optimal $\tilde{\mathbf{y}}$, $\mathbf{P2}$ can be reduced to
\begin{align}\label{eq:opt_3}
	(\mathbf{P3}):\ & \mathop{\max}_{\tilde{\mathbf{t}}, \tilde{\mathbf{r}}, \mathbf{W}_{\mathrm{T}}, \mathbf{w}_{\mathrm{L}},  \mathbf{v}_{\mathrm{A}}} \ \mathop{\min}_{\Delta} \quad \log_{2}\left(1+\gamma_{\mathrm{A}}\right) \\
	& \qquad \ \ \text{s.t. } \eqref{cons:ap_size}-\eqref{cons:uav_beam_unit}, \eqref{cons:min_user_rate_2}. \notag
\end{align}

Note that $\mathbf{P3}$ contains sum-of-ratio terms in the QoS constraints. To address this, we apply the quadratic transformation \cite{fp_skm_tsp}, which yields the following reformulated problem
\begin{subequations}
	\label{eq:opt_4}
	\begin{align}
		(\mathbf{P4}):\ & \mathop{\max}_{\tilde{\mathbf{t}}, \tilde{\mathbf{r}}, \mathbf{W}_{\mathrm{T}}, \mathbf{w}_{\mathrm{L}},  \mathbf{v}_{\mathrm{A}}, \mathbf{z} } \ \mathop{\min}_{\Delta} \quad \log_{2}\left(1+\gamma_{\mathrm{A}}\right) \\
		& \qquad \ \ \text{s.t. } R''_{k} \geq R_{k,\mathrm{min}}, \  \forall k \label{cons:min_user_rate_3} \\
		& \qquad \ \ \phantom{\text{s.t. }} \eqref{cons:ap_size}-\eqref{cons:uav_beam_unit} \notag
	\end{align}
\end{subequations}
where $R''_{k}\! = \! \log_{2}\left(1\!+\!\tilde{y}_{k}\right) \!-\! \tilde{y}_{k} 
\!+\! 2z_{k}\sqrt{1\!+\!\tilde{y}_{k}}\re\big\{\mathbf{h}^{\mathrm{H}}_{\mathrm{T},k}\mathbf{w}_{\mathrm{T},k}\big\} \!-\! z^{2}_k\Gamma_{k}$. 
In $\mathbf{P4}$, we introduce $\mathbf{z} = \left[z_{1}, z_{2}, \cdots, z_{K}\right]^{\mathrm{T}}$ to transform the original problem to a quadratic programming problem. More specifically, the optimal $z_{k}$ for $\mathbf{P4}$ is given by
\begin{equation}\label{eq:opt_z}
	z_{k} = {\sqrt{1+\tilde{y}_{k}}\re\left\{\mathbf{h}^{\mathrm{H}}_{\mathrm{T},k}\mathbf{w}_{\mathrm{T},k}\right\} }\Gamma_{k}^{-1}.
\end{equation}
Then, in terms of fixed $\mathbf{z}$, $\mathbf{P4}$ can be further reduced to
\begin{align}\label{eq:opt_5}
	(\mathbf{P5}):\ & \mathop{\max}_{\tilde{\mathbf{t}}, \tilde{\mathbf{r}}, \mathbf{W}_{\mathrm{T}}, \mathbf{w}_{\mathrm{L}},  \mathbf{v}_{\mathrm{A}}} \ \mathop{\min}_{\Delta} \quad \log_{2}\left(1+\gamma_{\mathrm{A}}\right) \\
	& \qquad \ \ \text{s.t. } \eqref{cons:ap_size}-\eqref{cons:uav_beam_unit}, \eqref{cons:min_user_rate_3}. \notag
\end{align}

Nevertheless, $\mathbf{P5}$ remains challenging to solve due to the tightly coupled variables and the infinitely non-convexity of the objective function. To address this, we propose a BCD-based algorithm that iteratively optimizes each variable while keeping the others fixed.
\vspace{-4mm}
\subsection{LBS Beamforming Design}
Firstly, we focus on optimizing the LBS beamforming $\mathbf{w}_{\mathrm{L}}$ under the maximum power constraint and QoS constraints. By defining $\bar{\mathbf{H}}_{\mathrm{LA}} = \mathbf{H}^{\mathrm{H}}_{\mathrm{LA}}\mathbf{v}_{\mathrm{A}}\mathbf{v}^{\mathrm{H}}_{\mathrm{A}}\mathbf{H}_{\mathrm{LA}}$ and $\mathbf{W}_{\mathrm{L}} = \mathbf{w}_{\mathrm{L}}\mathbf{w}^{\mathrm{H}}_{\mathrm{L}}$ with an implicit constraint $\rank({\mathbf{W}_{\mathrm{L}}}) = 1$, $\mathbf{P5}$ can be simplified as
\begin{subequations}
	\begin{align}
		(\mathbf{P6}):\quad \mathop{\min}_{\mathbf{W}_{\mathrm{L}}} \ &  -\tr({\bar{\mathbf{H}}_{\mathrm{LA}}\mathbf{W}_{\mathrm{L}}}) \\
		\text{s.t.}
		\ & \tr(\mathbf{W}_{\mathrm{L}}) \leq P_{\mathrm{L,max}} \\
		\ & \tr(\mathbf{\Phi}_{k}\mathbf{W}_{\mathrm{L}}) \leq \tilde{R}_{k},   \quad  \forall k \label{cons:lbs_min_rate}\\ 
		\ & \mathbf{W}_{\mathrm{L}} \succeq \mathbf{0} \label{cons:succ}
	\end{align}
\end{subequations}
where 
$
	\mathbf{\Phi}_{k} = z^2_{k}\mathbf{h}_{\mathrm{L},k}\mathbf{h}^{\mathrm{H}}_{\mathrm{L},k}
$, and 
$
\tilde{R}_{k} = \log_2(1+\tilde{y}_{k}) - \tilde{y}_{k} + 2z_{k}\sqrt{1+\tilde{y}_{k}}\mathrm{Re}\{ \mathbf{h}^{\mathrm{H}}_{\mathrm{T},k}\mathbf{w}_{\mathrm{T},k}\} - z^{2}_{k}\left(\sum_{i=1}^{K}\lvert \mathbf{h}^{\mathrm{H}}_{\mathrm{T},k}\mathbf{w}_{\mathrm{T},i} \lvert^{2} + \sigma^{2}_{\mathrm{T},k}\right) - R_{k,\mathrm{min}} \notag
$.
It is found that the QoS constraints \eqref{cons:lbs_min_rate} prevent us from deriving the semi-closed-form solution of $\mathbf{P6}$. Therefore, we propose the following proposition to transform $\mathbf{P6}$ into a tractable one.
\begin{proposition}\label{prop:lbs_beam}
	$\mathbf{P6}$ can be equivalently simplified to
	\begin{subequations}
		\begin{align}
			(\mathbf{P7}):\quad \mathop{\min}_{\mathbf{W}_{\mathrm{L}}} \ &  -\tr({\bar{\mathbf{H}}_{\mathrm{LA}}\mathbf{W}_{\mathrm{L}}}) \\
			\text{s.t.}
			\ & \tr(\mathbf{\Phi}\mathbf{W}_{\mathrm{L}}) \leq \bar{R} \\ 
			\ & \eqref{cons:succ} \notag
		\end{align}
	\end{subequations}
	where $\mathbf{\Phi} = \nu_{0}\mathbf{I}_{M} + \sum_{i=1}^{K}\nu_{i}\mathbf{\Phi}_{i}, \bar{R} = P_{\mathrm{L,max}} + \sum_{i=1}^{K}\tilde{R}_{i}$ and $\nu_{i} = \xi_{i}\bar{R} / (\xi_{0}P_{\mathrm{L,max}}+\sum_{j=1}^{K}\xi_{j}\tilde{R}_{j}), \ 0 \leq i \leq K$. Specifically, $\xi_{i}$ is the optimal dual variable associated with $\mathbf{P6}$, which can be obtained by solving the dual problem or subgradient method.
\end{proposition}
\begin{proof}
	Please refer to \cite{syf_ris_aj_1}.
\end{proof}
\vspace{-1mm}
By using Proposition \ref{prop:lbs_beam} and letting $\mathbf{W}_{\mathrm{L}} = \mathbf{\Phi}^{-\frac{1}{2}}\widetilde{\mathbf{W}}_{\mathrm{L}}\mathbf{\Phi}^{-\frac{1}{2}}$, $\mathbf{P6}$ can be equivalently rewritten as a classical MIMO capacity maximization problem, i.e.,
\begin{subequations}
	\begin{align}
		(\mathbf{P8}):\quad \mathop{\min}_{\widetilde{\mathbf{W}}_{\mathrm{L}}} \ &  \tr({\bar{\mathbf{H}}_{\mathrm{LA}}\mathbf{\Phi}^{-\frac{1}{2}}\widetilde{\mathbf{W}}_{\mathrm{L}}\mathbf{\Phi}^{-\frac{1}{2}}}) \\
		\text{s.t.}
		\ & \tr(\widetilde{\mathbf{W}}_{\mathrm{L}}) \leq \bar{R} \\ 
		\ & \eqref{cons:succ}. \notag
	\end{align}
\end{subequations}

It is well known that the optimal $\widetilde{\mathbf{W}}_{\mathrm{L}}$ for the classical MIMO capacity maximization problem is
\begin{equation}
	\widetilde{\mathbf{W}}_{\mathrm{L}} \! = \! \bar{R}\mathbf{u}\Big(\mathbf{\Phi}^{-\frac{1}{2}}{\bar{\mathbf{H}}_{\mathrm{LA}}}\mathbf{\Phi}^{-\frac{1}{2}}\Big)\mathbf{u}^{\mathrm{H}}\Big(\mathbf{\Phi}^{-\frac{1}{2}}{\bar{\mathbf{H}}_{\mathrm{LA}}}\mathbf{\Phi}^{-\frac{1}{2}}\Big)
\end{equation}
where $\mathbf{u}\left(\mathbf{\Phi}^{-\frac{1}{2}}{\bar{\mathbf{H}}_{\mathrm{LA}}}\mathbf{\Phi}^{-\frac{1}{2}}\right)$ is the eigenvector corresponding  to the largest eigenvalue of $\mathbf{\Phi}^{-\frac{1}{2}}{\bar{\mathbf{H}}_{\mathrm{LA}}}\mathbf{\Phi}^{-\frac{1}{2}}$. Therefore, by aligning the optimal $\widetilde{\mathbf{W}}_{\mathrm{L}}$ with the largest eigenvalues of $\mathbf{\Phi}^{-\frac{1}{2}}{\bar{\mathbf{H}}_{\mathrm{LA}}}\mathbf{\Phi}^{-\frac{1}{2}}$, the optimal closed-form solution of $\mathbf{w}_{\mathrm{L}}$ can be derived by
\begin{equation}\label{eq:opt_lbs_beam}
	\mathbf{w}_{\mathrm{L}} = \sqrt{\bar{R}}\mathbf{u}\Big(\mathbf{\Phi}^{-\frac{1}{2}}{\bar{\mathbf{H}}_{\mathrm{LA}}}\mathbf{\Phi}^{-\frac{1}{2}}\Big).
\end{equation}
It is noteworthy that the implicit rank-one constraint of $\mathbf{P6}$ can be satisfied by \eqref{eq:opt_lbs_beam}.
\vspace{-5mm}
\subsection{TBS Beamforming Design}
In this subsection, the optimization of the TBS beamforming  $\mathbf{W}_{\mathrm{T}}$ is investigated, whose corresponding subproblem can be formulated as
\begin{equation}
	\begin{aligned}
		(\mathbf{P9}):\quad \mathop{\min}_{\mathbf{W}_{\mathrm{T}}} \ &  \sum_{k=1}^{K} \lvert \mathbf{v}^{\mathrm{H}}_{\mathrm{A}}\mathbf{H}_{\mathrm{TA}}\mathbf{w}_{\mathrm{T},k} \lvert^{2} \ \text{s.t.}
		\ \eqref{cons:min_user_rate_3}.
%
	\end{aligned}
\end{equation}

It can be observed that $\mathbf{P9}$ is actually an inhomogeneous and separable quadratic constrained quadratic programming problem, which can be solved by convex optimization toolboxes like CVX in MATLAB. Nevertheless, the complexity of CVX is unbearable due to the multiple iterations. We first adopt the Lagrange multiplier-based method to handle the QoS constraints, i.e.,
\begin{equation}\label{porp:opt9}
		\mathop{\min}_{{\mathbf{W}}_{\mathrm{T}}} \ \mathcal{L} = \sum_{k=1}^{K} \lvert \mathbf{v}^{\mathrm{H}}_{\mathrm{A}}\mathbf{H}_{\mathrm{TA}}\mathbf{w}_{\mathrm{T},k} \lvert^{2}
		 - \sum_{k=1}^{K} \omega_{k}\left(\mathcal{R}''_{k} - R_{k,\mathrm{min}}\right)
\end{equation}
where $\left\{\omega_{k}\right\}^{K}_{k=1}$ are the multiple non-negative Lagrange multipliers. Then, armed with \eqref{porp:opt9}, we have the following proposition for solving $\mathbf{P9}$ in a semi-closed-form solution. 
\par
\begin{proposition}\label{prop:opt_gbs_beam}
	 The stationary point with the minimum transmit power of the problem $\mathbf{P9}$ is given by
	 \begin{equation}\label{eq:opt_gbs_beam} 
	 	\mathbf{w}_{\mathrm{T},k} = \omega_{k}z_{k}\sqrt{1+\tilde{y}_{k}}\mathbf{B}^{\dagger}\mathbf{h}_{\mathrm{T},k}
	 \end{equation}
	 where $\mathbf{B} \!=\! \mathbf{H}^{\mathrm{H}}_{\mathrm{TA}}\mathbf{v}_{\mathrm{A}}\mathbf{v}^{\mathrm{H}}_{\mathrm{A}}\mathbf{H}_{\mathrm{TA}} + \sum_{i=1}^{K}\omega_{i}z^{2}_{i}\mathbf{h}_{\mathrm{T},i}\mathbf{h}^{\mathrm{H}}_{\mathrm{T},i} \in \mathbb{C}^{N_{\mathrm{T}} \times N_{\mathrm{T}}}$.
\end{proposition}
\begin{proof}
	By taking the first-order derivative of $\mathcal{L}$ over the beamforming vectors (i.e. $\left\{\mathbf{w}_{\mathrm{T},k}\right\}^{K}_{k=1}$) and setting it to zero, we have
	\begin{equation}\label{eq:stationarity}
		\mathbf{B}\mathbf{w}_{\mathrm{T},k} = \omega_{k}z_{k}\sqrt{1+\tilde{y}_{k}} \mathbf{h}_{\mathrm{T},k}.
	\end{equation}
	Since $\mathbf{B}$ is positive semi-definite but generally rank-deficient (when $N_{\mathrm{T}} > K+1$), it is not invertible. However, the right-hand side vector $\omega_{k}z_{k}\sqrt{1+\tilde{y}_{k}} \mathbf{h}_{\mathrm{T},k}$ lies in the range space of $\mathbf{B}$, because $\mathbf{h}_{\mathrm{T},k}$ is one of the basis vectors spanning $\mathrm{Range}(\mathbf{B})$. Therefore, a solution to \eqref{eq:stationarity} exists. 
	Among all possible solutions, the one with the minimum Euclidean norm is given by
	\begin{equation}
		\mathbf{w}_{\mathrm{T},k} = \omega_{k}z_{k}\sqrt{1+\tilde{y}_{k}} \cdot \mathbf{B}^{\dagger} \mathbf{h}_{\mathrm{T},k}
	\end{equation}
	which corresponds to the minimum transmit power solution, akin to zero-forcing beamforming. Hence, the proof of Proposition \ref{prop:opt_gbs_beam} is completed.
\end{proof}
\begin{remark}
	It can be found that the beamforming vectors for terrestrial users are linear transformations of their corresponding channel vectors from Proposition \ref{prop:opt_gbs_beam}, where $\omega_{k}z_{k}\sqrt{1+y_{k}}$ is the linear power allocation coefficient, $\omega_{i}z^{2}_{i}$ in $\mathbf{B}$ is the priority coefficient for interference suppression, and $\mathbf{B}^{\dagger} \mathbf{h}_{\mathrm{T},k}$ is the beamforming direction. 
\end{remark}
It can be found that the QoS constraint must hold with equality for the optimal solution of $\mathbf{P9}$ according to Karush-Kuhn-Tucker (KKT) condition. As a result, we can determine the optimal $\left\{\omega^{\star}_{k}\right\}^{K}_{k=1}$ by the fixed-point algorithm, where $\omega^{\star}_{k}$ denotes the optimal dual variable. 
However, we can find that the overall complexity for calculating the beamforming vectors is $\mathcal{O}\left(KN^{2}_{\mathrm{T}}+N^{3}_{\mathrm{T}}\right)$. To address concern about the high computational complexity introduced by fixed point method, we consider setting a uniform $\omega_{k}$ for all terrestrial user for avoiding iteration. Generally, beamforming involves determining the beamforming direction and allocating power. As shown in \eqref{eq:opt_gbs_beam}, the beamforming direction is determined by the channel vectors and  $\left\{\omega_{k}\right\}^{K}_{k=1}$.  Consequently, the beamforming direction, denoted by $\left\{\widetilde{\mathbf{w}}_{\mathrm{T},k}\right\}^{K}_{k=1}$ is established. Then, we allocate the power by solving the following problem.
\begin{subequations}
	\begin{align}
		(\mathbf{P10}):\quad \mathop{\min}_{\mathbf{p}} \ &  \sum_{k=1}^{K} p_{k} \lvert \mathbf{v}^{\mathrm{H}}_{\mathrm{A}}\mathbf{H}_{\mathrm{TA}}\widetilde{\mathbf{w}}_{\mathrm{T},k} \lvert^{2} \\
		\text{s.t.}
		\ & \alpha_{1,k}p_{k} - \sum_{i \neq k}^{K}\alpha_{2,k,i}p_{i} \geq \beta_{k}, \quad \forall k \\ 
		\ & p_{k} \geq 0 ,\quad \forall k
	\end{align}
\end{subequations}
where $\mathbf{p} = \left[p_1, p_2, \cdots, p_{K}\right]^{\mathrm{T}}$ denotes the power allocation vector of terrestrial users, $\widetilde{\mathbf{w}}_{\mathrm{T},k} = \frac{\mathbf{B}^{\dagger}\mathbf{h}_{\mathrm{T},k}}{\Vert \mathbf{B}^{\dagger}\mathbf{h}_{\mathrm{T},k} \Vert}$, $\alpha_{1,k} =\lvert \mathbf{h}^{\mathrm{H}}_{\mathrm{T},k}\widetilde{\mathbf{w}}_{\mathrm{T},k} \lvert^{2}$, $\alpha_{2,k,i} = \left( 2^{R_{k,\mathrm{min}}} - 1\right)\lvert \mathbf{h}^{\mathrm{H}}_{\mathrm{T},k}\widetilde{\mathbf{w}}_{\mathrm{T},i} \rvert^{2}$, and $\beta_{k} = \left( 2^{R_{k,\mathrm{min}}} - 1\right)\Big(\lvert \mathbf{h}^{\mathrm{H}}_{\mathrm{L},k}\mathbf{w}_{\mathrm{L}} \lvert^{2} +\sigma^{2}_{\mathrm{T},k} \Big)$. 
The overall complexity of solving $\mathbf{P10}$ is given by $\mathcal{O}\left(K^{3.5}\right)$.
\vspace{-4mm}
\subsection{AO based Optimization Framework for $\mathbf{v}_{\mathrm{A}}, \tilde{\mathbf{r}}$}
In this subsection, we aim to optimize the receive combiner $\mathbf{v}_{\mathrm{A}}$ and receive APV $\tilde{\mathbf{r}}$ under the imperfect angular CSI $\Delta$. The corresponding subproblem can be formulated as
\begin{align}\label{eq:opt_11}
	(\mathbf{P11}):\ & \mathop{\max}_{\tilde{\mathbf{r}}, \mathbf{v}_{\mathrm{A}}} \ \mathop{\min}_{\Delta} \quad \log_{2}\left(1+\gamma_{\mathrm{A}}\right) \\	
	& \ \ \text{s.t. } \eqref{cons:ap_size}, \eqref{cons:max_ap_uav}, \eqref{cons:uav_beam_unit}. \notag
\end{align}

Note that $\mathbf{P11}$ is hard to solve directly due to the non-convex constraint $\eqref{cons:max_ap_uav}$ and the imperfect angular CSI $\Delta$. To handle this problem, we first employ a convex hull-based approach to handle the CSI imperfection $\Delta$\footnote{The concept of convex hull implies that any channel matrix in the uncertainty set is a weighted combination of discrete samples \cite{boyd2004}.}, and then obtain the closed-form solution of $\mathbf{v}_{\mathrm{L}}$ by applying the linear minimum-mean-square-error (MMSE) combiner. Finally, we propose an efficient algorithm based on the geometric boundary method to design the APV.
\subsubsection{\textbf{Minimization Problem Over Imperfect Angular CSI}}
Due to the fact that the minimization problem related to $\Delta$ is irrelevant to the constraints \eqref{cons:ap_size}, \eqref{cons:max_ap_uav} and \eqref{cons:uav_beam_unit}, we consider the worst-case problem as
\begin{equation}
	(\mathbf{P12}):\ \mathop{\max}_{\Delta} \quad \mathbf{v}^{\mathrm{H}}_{\mathrm{A}}\mathbf{H}_{\mathrm{JA}}\mathbf{w}_{\mathrm{J}}\mathbf{w}^{\mathrm{H}}_{\mathrm{J}}\mathbf{H}^{\mathrm{H}}_{\mathrm{JA}}\mathbf{v}_{\mathrm{A}}.
\end{equation}

Note $\mathbf{P12}$ is non-concave with respect to $\Delta$ due to the nonlinear objective function, the intractable angular range, and the unknown jamming beamforming $\mathbf{w}_{\mathrm{J}}$. 
To ensure $\mathbf{P12}$ feasibility, we utilize the Cauchy-Schwarz inequality and the convex hull-based approach to tackle the unknown $\mathbf{w}_{\mathrm{J}}$ and $\Delta$, respectively. 
First of all, by using the Cauchy-Schwarz inequality, we can obtain the upper bound of received jamming power, which is written as
\begin{equation}\label{eq:Cauchy_ineq}
	\mathbf{v}^{\mathrm{H}}_{\mathrm{A}}\mathbf{H}_{\mathrm{JA}}\mathbf{w}_{\mathrm{J}}\mathbf{w}^{\mathrm{H}}_{\mathrm{J}}\mathbf{H}^{\mathrm{H}}_{\mathrm{JA}}\mathbf{v}_{\mathrm{A}} \leq \hat{p}_{\mathrm{J}}\mathbf{v}^{\mathrm{H}}_{\mathrm{A}}\mathbf{H}_{\mathrm{JA}}\mathbf{H}^{\mathrm{H}}_{\mathrm{JA}}\mathbf{v}_{\mathrm{A}}
\end{equation}
where $\hat{p}_{\mathrm{J}}$ denotes the estimation of the jammer’s transmit power, which can be obtained by the rotational invariance techniques. Then, we turn to handle the CSI imperfection $\Delta$ by leveraging the convex hull property. We first uniformly discretize all the angles inside $\Delta$, i.e., 
\begin{equation}
	\begin{aligned}
		\theta^{(p)} &= \theta_L + (i_1-1) \Delta \theta, \ i_1 = 1,\ldots,Q_1 \\
		\phi^{(p)} &= \phi_L + (i_2-1) \Delta \phi, \ i_2 = 1,\ldots,Q_2
	\end{aligned}
\end{equation}
where $p = (i_2-1)Q_1 + i_1$ indexes each grid point, $\Delta \theta = (\theta_U - \theta_L)/(Q_1 - 1)$, and $\Delta \phi = (\phi_U - \phi_L)/(Q_2 - 1)$. The $p$-th channel realization is $\mathbf{H}_{\mathrm{JA}}^{(p)} = \mathbf{F}_{\mathrm{JA}}^{(p),\mathrm{H}}(\tilde{\mathbf{r}}) \bm{\Sigma}_{\mathrm{JA}} \mathbf{G}_{\mathrm{JA}}$, where $\mathbf{F}_{\mathrm{JA}}^{(p),\mathrm{H}}(\tilde{\mathbf{r}})= [\mathbf{f}_{\mathrm{JA}}^{(p)}(\mathbf{r}_1), \ldots, \mathbf{f}_{\mathrm{JA}}^{(p)}(\mathbf{r}_{N_{\mathrm{A}}})]$ incorporates the angles $\!\{\theta^{(p)}, \phi^{(p)}\}$, with $\mathbf{f}_{\mathrm{JA}}^{(p)}(\mathbf{r}_m) \!\!= \big[ e^{j \frac{2\pi}{\lambda} \mathbf{r}_m^{\mathrm{T}} \mathbf{n}^{\mathrm{r},(p)}_{\mathrm{JA},\tilde{l}}} \big]_{1 \leq \tilde{l} \leq L^{\mathrm{r}}_{\mathrm{JA}}}^{\mathrm{T}}$, and $\mathbf{n}^{\mathrm{r},(p)}_{\mathrm{JA},\tilde{l}} = \big[ \cos \theta_{\mathrm{JA},{\tilde{l}}}^{\mathrm{r},(p)} \cos \phi_{\mathrm{JA},{\tilde{l}}}^{\mathrm{r},(p)}, \cos \theta_{\mathrm{JA},{\tilde{l}}}^{\mathrm{r},(p)} \sin \phi_{\mathrm{JA},{\tilde{l}}}^{\mathrm{r},(p)} \big]^{\mathrm{T}}$. Then we can construct a convex hull based on all discrete angles as
\begin{equation}\label{eq:conv_hall}
	\mathcal{S} = \Bigg\{ \sum_{p=1}^{Q} \mu_p \mathbf{H}_{\mathrm{JA}}^{(p)} \mathbf{H}_{\mathrm{JA}}^{(p),\mathrm{H}} \bigg| \sum_{p=1}^{Q} \mu_p = 1, \mu_p \geq 0 \Bigg\}
\end{equation}
where $Q = Q_1 Q_2$ and $\mu_p \geq 0$ are weights summing to one for the $p$-th channel realization.
Substituting \eqref{eq:conv_hall} into $\mathbf{P12}$, the intractable $\mathbf{P12}$ can be equivalently reformulated as
\begin{equation}
	\max_{\mathcal{S}} \quad \mathbf{v}_{\mathrm{A}}^{\mathrm{H}} \sum_{p=1}^{Q} \mu_p \mathbf{H}_{\mathrm{JA}}^{(p)} \mathbf{H}_{\mathrm{JA}}^{(p),\mathrm{H}} \mathbf{v}_{\mathrm{A}},
\end{equation}
which is equivalent to
\begin{subequations}
	\begin{align}
		(\mathbf{P13}): \max_{\bm{\mu}} \ & \mathbf{v}_{\mathrm{A}}^{\mathrm{H}} \sum_{p=1}^{Q} \mu_p \mathbf{H}_{\mathrm{JA}}^{(p)} \mathbf{H}_{\mathrm{JA}}^{(p),\mathrm{H}} \mathbf{v}_{\mathrm{A}} \label{eq:obj_p13} \\
		\text{s.t.} \ & \sum_{p=1}^{Q} \mu_p = 1, \mu_p \geq 0
	\end{align}
\end{subequations}
where $\bm{\mu} = [\mu_1, \mu_2, \ldots, \mu_{Q}]^{\mathrm{T}} \in \mathbb{R}^{Q \times 1}$. Then, we have the following proposition to choose $\bm{\mu}$, such that the worst case is achieved.
\begin{proposition}\label{prop:mu_opt}
	For given $\mathbf{v}_{\mathrm{A}}$ and $\tilde{\mathbf{r}}$, the optimal $\mu_p$ is:
	\begin{equation}\label{eq:opt_mu}
		\mu_p^{\star} = \frac{\mathbf{v}_{\mathrm{A}}^{\mathrm{H}} \mathbf{H}_{\mathrm{JA}}^{(p)} \mathbf{H}_{\mathrm{JA}}^{(p),\mathrm{H}} \mathbf{v}_{\mathrm{A}}}{\sum_{p'=1}^{Q} \mathbf{v}_{\mathrm{A}}^{\mathrm{H}} \mathbf{H}_{\mathrm{JA}}^{(p')} \mathbf{H}_{\mathrm{JA}}^{(p'),\mathrm{H}} \mathbf{v}_{\mathrm{A}}}.
	\end{equation}
\end{proposition}
\begin{proof}
	Please refer to \cite{syf_convex_hull}.
\end{proof}
Based on the relationship between $\mu_{p}$, $\mathbf{v}_{\mathrm{A}}$ and $\tilde{\mathbf{r}}$ in Proposition \ref{prop:mu_opt}, we can alternatively optimize $\mu_{p}$, $\mathbf{v}_{\mathrm{A}}$ and $\tilde{\mathbf{r}}$ to obtain the suboptimal solution to the max-min problem \eqref{eq:opt_11}.
\subsubsection{\textbf{Maximization Problem w.r.t $\mathbf{v}_{\mathrm{A}}$}}
In this subproblem, we investigate the design of $\mathbf{v}_{\mathrm{A}}$ for maximizing the receive SINR. Given $\bm{\mu}$ and $\tilde{\mathbf{r}}$, the outer maximization problem w.r.t $\mathbf{v}_{\mathrm{A}}$ is given by
\begin{align}\label{eq:opt_14}
	(\mathbf{P14}):\ & \mathop{\max}_{\mathbf{v}_{\mathrm{A}}} \ \log_{2}\left(1+\gamma'_{\mathrm{A}}\right) \ \text{s.t. } \eqref{cons:uav_beam_unit} 
\end{align}
where $
\gamma'_{\mathrm{A}} = { \mathbf{v}^{\mathrm{H}}_{\mathrm{A}}\mathbf{H}_{\mathrm{LA}}\mathbf{w}_{\mathrm{L}}\mathbf{w}^{\mathrm{H}}_{\mathrm{L}}\mathbf{H}^{\mathrm{H}}_{\mathrm{LA}}\mathbf{v}_{\mathrm{A}} } \big(\mathbf{v}^{\mathrm{H}}_{\mathrm{A}}\widetilde{\mathbf{A}}\mathbf{v}_{\mathrm{A}} \big)^{-1}
$, 
and $\widetilde{\mathbf{A}}$ is defined as
\begin{equation}\label{eq:tilde_A}
	\widetilde{\mathbf{A}} = \mathbf{A}_{1} + \hat{p}_{\mathrm{J}}\sum_{p=1}^{Q} \mu_{p} \mathbf{H}^{(p)}_{\mathrm{JA}} \mathbf{H}^{(p),\mathrm{H}}_{\mathrm{JA}} \in \mathbb{C}^{N_{\mathrm{A}} \times N_{\mathrm{A}}}.
\end{equation}
As is known to all, the linear MMSE combiner is the optimal combiner for maximizing the receive SINR, which can balance the interference and noise at the receiver. Thus, we directly adopt MMSE combiner for $\mathbf{v}_{\mathrm{A}}$, as given by
\begin{equation}\label{eq:opt_v_A}
	\mathbf{v}_{\mathrm{A}} = \frac{\left(\mathbf{H}_{\mathrm{LA}}\mathbf{w}_{\mathrm{L}}\mathbf{w}^{\mathrm{H}}_{\mathrm{L}}\mathbf{H}^{\mathrm{H}}_{\mathrm{LA}} + \widetilde{\mathbf{A}}\right)^{-1}\mathbf{H}_{\mathrm{LA}}\mathbf{w}_{\mathrm{L}}}{\Big \Vert \left(\mathbf{H}_{\mathrm{LA}}\mathbf{w}_{\mathrm{L}}\mathbf{w}^{\mathrm{H}}_{\mathrm{L}}\mathbf{H}^{\mathrm{H}}_{\mathrm{LA}} + \widetilde{\mathbf{A}}\right)^{-1}\mathbf{H}_{\mathrm{LA}}\mathbf{w}_{\mathrm{L}} \Big \Vert}.
\end{equation}
\subsubsection{\textbf{Maximization Problem w.r.t $\tilde{\mathbf{r}}$}}
After optimizing $\bm{\mu}$ and $\mathbf{v}_{\mathrm{A}}$, we turn to the design of receive APV $\tilde{\mathbf{r}}$, whose corresponding subproblem can be formulated as
\begin{align}\label{eq:opt_15}
	(\mathbf{P15}):\ & \mathop{\max}_{\tilde{\mathbf{r}}} \  \frac{\lvert \mathbf{v}^{\mathrm{H}}_{\mathrm{A}} \mathbf{H}_{\mathrm{LA}}\left(\tilde{\mathbf{r}}\right)\mathbf{w}_{\mathrm{L}} \lvert^2}{\mathbf{v}^{\mathrm{H}}_{\mathrm{A}}\widetilde{\mathbf{A}}\left(\tilde{\mathbf{r}}\right)\mathbf{v}_{\mathrm{A}}} \ \text{s.t. } \eqref{cons:ap_size}, \eqref{cons:max_ap_uav}.
\end{align}

$\mathbf{P15}$ is challenging to handle due to the fractional form of objective function. To address this issue, we first adopt the
Dinkelbach’s method to transform \eqref{eq:opt_15} into an equivalent form, i.e.,
\begin{align}\label{eq:opt_16}
	\mathop{\min}_{\tilde{\mathbf{r}}} \ & f_{1} = \kappa\mathbf{v}^{\mathrm{H}}_{\mathrm{A}}\widetilde{\mathbf{A}}\left(\tilde{\mathbf{r}}\right)\mathbf{v}_{\mathrm{A}} - \lvert \mathbf{v}^{\mathrm{H}}_{\mathrm{A}}\mathbf{H}_{\mathrm{LA}}\left({\tilde{\mathbf{r}}_{\mathrm{A}}}\right)\mathbf{w}_{\mathrm{L}} \rvert^{2} \\
	\text{s.t.}
	\ & \eqref{cons:ap_size}, \eqref{cons:max_ap_uav}. \notag
\end{align}
To deal with problem \eqref{eq:opt_16}, we adopt a cyclic coordinate descent (CCD) framework but with a non-convex constraint to find at least a locally optimal solution. Instead of directly solving problem \eqref{eq:opt_16}, we will successively optimize upper bounds of $f_{1}$ in an antenna-by-antenna manner. As such, we can obtain $N_\mathrm{A} + 1$ subproblems and update them by the following CCD algorithm, i.e., 
\begin{equation}
	\left\{
	\begin{aligned}
		& \kappa^{(t+1)} \!=\! f_{\kappa}\Big(\kappa, \mathbf{r}^{(t)}_{1}, \mathbf{r}^{(t)}_{2}, \cdots, \mathbf{r}^{(t)}_{N_{\mathrm{A}}}\Big) \\
		& \mathbf{r}^{(t+1)}_{1} \!=\! \mathop{\arg\min}\limits_{\mathbf{r}_{1} \in \Omega_{1}} f_{1}\Big(\kappa^{(t+1)}, \mathbf{r}_{1}, \mathbf{r}^{(t)}_{2}, \cdots, \mathbf{r}^{(t)}_{N_{\mathrm{A}}}\Big) \\
		& \qquad \qquad \qquad \qquad \vdots \\
		& \mathbf{r}^{(t+1)}_{N_{\mathrm{A}}} \!=\! \mathop{\arg\min}\limits_{\mathbf{r}_{N_{\mathrm{A}}} \in \Omega_{N_{\mathrm{A}}}} f_{1}\Big(\kappa^{(t+1)}, \mathbf{r}^{(t+1)}_{1}, \mathbf{r}^{(t+1)}_{2}, \cdots, \mathbf{r}_{N_{\mathrm{A}}}\Big)
	\end{aligned}
	\right.
\end{equation}
where $f_{\kappa}$ is the mapping function from $\left(\kappa, \mathbf{r}^{(t)}_{1}, \mathbf{r}^{(t)}_{2}, \cdots, \mathbf{r}^{(t)}_{N_{\mathrm{A}}}\right)$ to closed-form solution, and $\Omega_{m} \!\!=\!\! \left\{\mathbf{r}_{m} \middle| \left(\Vert \mathbf{r}_{m} - \mathbf{r}_{i} \Vert_{2} \geq D, i \neq m \right)\cap \left(\mathbf{r}_{m} \in \mathcal{C}_{\mathrm{r}} \right) \right\}$ represents the feasible region for optimizing $\mathbf{r}_{m}$ at the $(t+1)$-th iteration. Using the above procedure, the subproblem w.r.t. $\kappa$ admits the following solution, i.e.,
\begin{equation}\label{eq:opt_kappa}
	\kappa^{(t+1)} = {\big\lvert \mathbf{v}^{\mathrm{H}}_{\mathrm{A}}\mathbf{H}_{\mathrm{LA}}\big({\tilde{\mathbf{r}}^{(t)}}\big)\mathbf{w}_{\mathrm{L}} \big\rvert^{2}}\Big(\mathbf{v}^{\mathrm{H}}_{\mathrm{A}}\widetilde{\mathbf{A}}\big(\tilde{\mathbf{r}}^{(t)}\big)\mathbf{v}_{\mathrm{A}}\Big)^{-1}.
\end{equation}

Now, assuming $\kappa^{(t+1)}$ for variable $\kappa$ , we turn to solve the subproblem w.r.t. $\mathbf{r}_{m}$. Denote the gradient vector and Hessian matrix of $f_{1}$ over $\mathbf{r}_{m}$ by $\nabla f_{1}\left(\mathbf{r}_{m}\right) \in \mathbb{R}^{2 \times 1}$ and $\nabla^{2} f_{1}\left(\mathbf{r}_{m}\right) \in \mathbb{R}^{2 \times 2}$, respectively, with their derivations provided in Appendix \ref{app:A}. Then, we construct a positive real number $\tau_{m}$ making $\tau_{m}\mathbf{I}_{2} \succeq \nabla^{2} f_{1}\left(\mathbf{r}_{m}\right)$, ensured when $\tau_{m} \geq \lambda_{\max}\left(\nabla^{2} f_{1}\left(\mathbf{r}_{m}\right)\right)$ with $\lambda_{\max}(\cdot)$ denoting the maximum eigenvalue. Thus, based on Taylor’s theorem, we can find a quadratic surrogate function to globally upper-bound the objective function $f_{1}\left(\mathbf{r}_{m}\right)$ as
\begin{equation}\label{eq:uav_upper}
		f_{1}\left(\mathbf{r}_{m}\right) \! \leq \! f_{1}\big({\mathbf{r}}^{(t)}_{m}\big) \!+\! {\nabla}f^{\mathrm{T}}_{1}\big({\mathbf{r}}^{(t)}_{m}\big)\big({\mathbf{r}_{m}} \!-\! {\mathbf{r}^{(t)}_{m}}\big) 
			 + \frac{1}{2}\tau^{(t)}_{m}\Vert{\mathbf{r}_{m}} \!-\! {\mathbf{r}^{(t)}_{m}} \Vert^{2} .
\end{equation}
With the upper bound in \eqref{eq:uav_upper}, the problem can be recast as
\begin{align}\label{eq:sca}
	 \mathop{\min}_{{\mathbf{r}_{m}}} \ & \frac{\tau^{(t)}_{m}}{2}\mathbf{r}^{\mathrm{T}}_{m}\mathbf{r}_{m} + \Big({\nabla}f_{1}\big({\mathbf{r}}^{(t)}_{m}\big) - \tau^{(t)}_{m}\mathbf{r}^{(t)}_{m}\Big)^{\mathrm{T}}\mathbf{r}_{m} \\
	\text{s.t.}
	\ & \eqref{cons:ap_size}, \eqref{cons:max_ap_uav}. \notag
\end{align} 
Then, the stationary point for minimizing \eqref{eq:sca} by neglecting constraints \eqref{cons:ap_size} and \eqref{cons:max_ap_uav} is obtained in closed form as
\begin{equation}
	 \mathbf{r}^{\star}_{m} = \mathbf{r}^{(t)}_{m} - {\nabla}f_{1}\big({\mathbf{r}}^{(t)}_{m}\big) / \tau^{(t)}_{m}.
\end{equation}
If $\mathbf{r}^{\star}_{m}$ satisfies the constraints \eqref{cons:ap_size} and \eqref{cons:max_ap_uav}, it is exactly the optimal solution. Otherwise, this stationary point is invalid. With the definition of the stationary point, we have the following definition and proposition for optimal solution of problem \eqref{eq:sca}, which states a critical property.
\begin{definition}
	We say the point $\mathbf{r}_{m'}$ satisfying $\Vert \mathbf{r}_{m'} \!-\! \mathbf{r}^{\star}_{m} \Vert \!\leq\! R$, is an active point with respect to $\mathbf{r}^{\star}_{m}$ within a region of radius $R$.
\end{definition}
\begin{proposition}\label{prop:bound}
	If the stationary point of problem \eqref{eq:sca} is invalid, the optimal solution of problem \eqref{eq:sca} must lie in the boundary of the feasible region $\Omega_{m}$.
\end{proposition}
\begin{proof}
	Please refer to Appendix \ref{app:B}.
\end{proof}

Based on Proposition~\ref{prop:bound}, we construct three categories of candidate points for antenna placement. These include:
\begin{enumerate}
	\item Candidate line intersections (CII): intersections between the $D$-circle centered at an active point and the straight line connecting it to the stationary point;
	\item Candidate circle intersections (CCI): intersections between the $D$-circles of any two active points;
	\item Candidate boundary intersections (CBI): intersections between an active point's $D$-circle and the boundary of the feasible moving area.
\end{enumerate}

Due to the high computational cost of evaluating all possible intersections among three types of points for each cluster center, we restrict our computation to only those intersections between the stationary point and active points lying within a radius of $
R \!=\! D \!+\! \big\Vert {\nabla}f^{\mathrm{T}}_{1}\big({\mathbf{r}}^{(t)}_{m}\big)/\tau^{(t)}_{m} \big\Vert
$.
To further reduce the computational complexity, we employ \textit{KD-Tree} to efficiently retrieve candidate active points. A \textit{KD-Tree} is a space-partitioning binary tree designed for organizing points in a $K$-dimensional space, which enables fast nearest neighbor and range searches. By recursively splitting the space along alternating axes, the \textit{KD-Tree} reduces the search complexity from $\mathcal{O}(N)$ to $\mathcal{O}(\log_{2}(N))$ on average for balanced trees. Subsequently, we generate candidate intersection points based solely on these active points and apply the \textit{KD-Tree} again to filter out those that do not satisfy all constraints. The final output is selected as the nearest feasible point. The complete procedure for finding $\mathbf{r}_{m}$ is outlined in Algorithm~\ref{alg:geometric_boundary}. Remarkably, compared with \cite{ARXIV_ZCJ}, our proposed method requires no iteration.
\vspace{-3mm}
\subsection{LBS APV Design}
Here, we focus on designing the transmit APV at the LBS, which can be formulated as
\begin{subequations}\label{eq:opt16}
	\begin{align}
		(\mathbf{P16}):\quad \mathop{\min}_{\tilde{\mathbf{t}}} \ & f_{2} = -\lvert \mathbf{v}^{\mathrm{H}}_{\mathrm{A}}\mathbf{H}_{\mathrm{LA}}\left({\tilde{\mathbf{t}}}\right)\mathbf{w}_{\mathrm{L}} \lvert^{2} \label{obj:prob16} \\
		\text{s.t.}
		\ & f_{3,k}\left({\tilde{\mathbf{t}}}\right) \leq -R_{k,\mathrm{min}},  \quad  \forall k\label{cons:P17_min_rate}\\
		\ & \eqref{cons:ap_size}, \eqref{cons:max_ap_lbs} \notag
	\end{align}
\end{subequations}
where $f_{3,k}\big({\tilde{\mathbf{t}}}\big) = -R''_{k}$. Similar to the optimization of $\tilde{\mathbf{r}}$, we also successively optimize the upper bounds of $f_{2}$ by constructing surrogate functions in an antenna-by-antenna manner.

The gradient vector and Hessian matrix of \( f_{2} \) and \( f_{3,k} \) with respect to \( \mathbf{t}_{n} \) are denoted as \( \nabla f_{2}\left(\mathbf{t}_{n}\right) \in \mathbb{R}^{2 \times 1} \), \( \nabla^{2} f_{2}\left(\mathbf{t}_{n}\right) \in \mathbb{R}^{2 \times 2} \), \( \nabla f_{3,k}\left(\mathbf{t}_{n}\right) \in \mathbb{R}^{2 \times 1} \), and \( \nabla^{2} f_{3,k}\left(\mathbf{t}_{n}\right) \in \mathbb{R}^{2 \times 2} \), respectively. Their derivations follow similar procedures to those in Appendix \ref{app:A} and are omitted here for brevity. Thus, the surrogate function can be expressed as
\begin{equation} %
		f_{2}\left(\mathbf{t}_{n}\right) \! \leq \!  f_{2}\big({\mathbf{t}}^{(t)}_{n}\big) + {\nabla}f^{\mathrm{T}}_{2}\big({\mathbf{t}}^{(t)}_{n}\big)\big({\mathbf{t}_{n}} - {\mathbf{t}^{(t)}_{n}}\big) 
		 + \frac{1}{2}\eta^{(t)}_{n}\Vert{\mathbf{t}_{n}} - {\mathbf{t}^{(t)}_{n}} \Vert^{2} 
\end{equation}
\begin{equation}
	\begin{aligned}
		 f_{3,k}\left(\mathbf{t}_{n}\right) & \leq  f_{3,k}\big({\mathbf{t}}^{(t)}_{n}\big) \!+\! {\nabla}f^{\mathrm{T}}_{3,k}\big({\mathbf{t}}^{(t)}_{n}\big)\big({\mathbf{t}_{n}}\! -\! {\mathbf{t}^{(t)}_{n}}\big) \\
		& \ \ \ + \frac{1}{2}\upsilon^{(t)}_{k,n}\Vert{\mathbf{t}_{n}} \!-\! {\mathbf{t}^{(t)}_{n}} \Vert^{2} \\
		& = \underbrace{\frac{\upsilon^{(t)}_{k,n}}{2}\mathbf{t}^{\mathrm{T}}_{n}\mathbf{t}_{n} + \Big({\nabla}f	_{3,k}\big({\mathbf{t}}^{(t)}_{n}\big) - \upsilon^{(t)}_{k,n}\mathbf{t}^{(t)}_{n}\Big)^{\mathrm{T}}\mathbf{t}_{n}}_{u_{3,k}\left(\mathbf{t}_{n}\right)} \\
		& \ \ \ +  \underbrace{f_{3,k}\big({\mathbf{t}}^{(t)}_{n}\big) - {\nabla}f^{\mathrm{T}}_{3,k}\big({\mathbf{t}}^{(t)}_{n}\big){\mathbf{t}^{(t)}_{n}} + \frac{\upsilon^{(t)}_{k,n}}{2}\mathbf{t}^{(t),\mathrm{T}}_{n}\mathbf{t}^{(t)}_{n}}_{u_{c,k}}
	\end{aligned}
\end{equation}
where $\eta_{n}$ is a positive number with $\eta_{n}\mathbf{I}_{2} \succeq \nabla^{2} f_{2}\left(\mathbf{t}_{n}\right)$, which can be achieved by setting $\eta_{n}$ greater than the largest eigenvalue of $\nabla^{2} f_{2}\left(\mathbf{t}_{n}\right)$. Similarly, let $\upsilon_{k,n}$ be a positive scalar such that $\upsilon_{k,n}\mathbf{I}_{2} \succeq \nabla^{2} f_{3,k}\left(\mathbf{t}_{n}\right)$, ensured when $\upsilon_{k,n} \geq \lambda_{\max}\left(\nabla^{2} f_{3,k}\left(\mathbf{t}_{n}\right)\right)$.
\begin{algorithm}[tb]
	\SetAlgoLined
	\SetKwInOut{Input}{Input}\SetKwInOut{Output}{Output}
	
	\caption{Geometric Boundary Algorithm for Antenna Position Optimization}
	\label{alg:geometric_boundary}
	
	\Input{$\mathbf{r}^{\star}_{m}$, $D$, $\tau_{m}$, ${\nabla}f_{1}\left({\mathbf{r}}_{m}\right)$
	}
	\Output{$\mathbf{r}_{m}$}
	
	\BlankLine
	
	\eIf{$\mathbf{r}^{\star}_{m}$ satisfies all constraints}{
		$\mathbf{r}_{m}$ = $\mathbf{r}^{\star}_{m}$.
	}{
		\textbf{Step 1: Compute Radius} 
		$R = D + \left\Vert {\nabla}f_{1}\left({\mathbf{r}}_{m}\right)/\tau_{m} \right\Vert$.
		
		\textbf{Step 2: Find Active Points Using \textit{KD-Tree}}
		Use \textit{KD-Tree} to find all active points within distance $R$ from $\mathbf{r}^{\star}_{m}$.
		
		\textbf{Step 3: Calculate Intersections}
		Calculate CII, CCI and CBI.
		
		\textbf{Step 4: Filter Candidate Points Using \textit{KD-Tree}}
		Select points that satisfy all constraint conditions and lie within the $\left\Vert {\nabla}f_{1}\left({\mathbf{r}}_{m}\right)/\tau_{m} \right\Vert$-radius circle centered at the stationary point. 
		
		\textbf{Step 5: Choose the Optimal Candidate Point}
	}
\end{algorithm}

With the upper bound, the optimization problem of the $n$-th transmit antenna position $\mathbf{t}_{n}$ at the $(t+1)$-th iteration can be relaxed as
\begin{subequations}\label{eq:prop_lbs_apv}
	\begin{align}
		\mathop{\min}_{{\mathbf{t}_{n}}} \ & \frac{\eta^{(t)}_{n}}{2}\mathbf{t}^{\mathrm{T}}_{n}\mathbf{t}_{n} + \Big({\nabla}f_{2}\big({\mathbf{t}}^{(t)}_{n}\big) - \eta^{(t)}_{n}\mathbf{t}^{(t)}_{n}\Big)^{\mathrm{T}}\mathbf{t}_{n} \\
		\text{s.t.} \ & u_{3,k}\left(\mathbf{t}_{n}\right)  \leq -R_{k,\mathrm{min}} - u_{c,k} \\
		\ & \eqref{cons:ap_size}, \eqref{cons:max_ap_uav}. \notag
	\end{align}
\end{subequations} 

The problem \eqref{eq:prop_lbs_apv} involves minimizing the objective function, which is a quadratic function of the variable \(\mathbf{t}_{n}\). Similarly, the constraint functions \(u_{3,k}\left(\mathbf{t}_{n}\right)\) for \(k = 1, \dots, K\) are also quadratic functions. Due to their quadratic nature, both objective function and \(u_{3,k}\left(\mathbf{t}_{n}\right)\) have well-defined minimum points, which can be derived as follows:
\begin{itemize}
	\item For the objective function, the minimum occurs at the stationary point \(\mathbf{t}_{n}^{\star} = \mathbf{t}^{(t)}_{n} - {\nabla}f_{2}\big({\mathbf{t}}^{(t)}_{n}\big) / \eta^{(t)}_{n}\).
	\item For each constraint function \(u_{3,k}\left(\mathbf{t}_{n}\right)\), the minimum occurs at \(\mathbf{t}_{n}^{k} = \mathbf{t}^{(t)}_{n} - {\nabla}f_{3,k}\big({\mathbf{t}}^{(t)}_{n}\big) / \upsilon^{(t)}_{k,n}\).
\end{itemize}
To address this, we propose the following solution strategy:
\begin{enumerate}
	\item \textbf{Case 1: Stationary Point Satisfies All Constraints}  
	
	If the point \(\mathbf{t}_{n}^{\star}\) satisfies all constraints, including \eqref{cons:P17_min_rate}, \eqref{cons:ap_size} and \eqref{cons:max_ap_lbs}, it is returned as the optimal solution.
	
	\item \textbf{Case 2: No Active Points Within Radius \(\big(D + \big\|{\nabla}f_{2}\big({\mathbf{t}}^{(t)}_{n}\big) / \eta^{(t)}_{n}\big\|\big)\), QoS Constraint Violated}  
	
	If there are no active points (i.e., other \(\mathbf{t}_{n'}\) for \(n' \neq n\)) within a radius \(\big(D + \big\|{\nabla}f_{2}\big({\mathbf{t}}^{(t)}_{n}\big) / \eta^{(t)}_{n}\big\|\big)\) centered at \(\mathbf{t}_{n}^{\star}\), and the QoS constraint \eqref{cons:P17_min_rate} is not satisfied, the problem is reformulated as follows
	\begin{subequations}\label{eq:prob_t}
		\begin{align}
			\mathop{\min}_{\mathbf{t}_{n}} \ & \|\mathbf{t}_{n} - \mathbf{t}_{n}^{\star}\|_2 \label{eq:obj_dist} \\
			\text{s.t.} \ & \|\mathbf{t}_{n} - \mathbf{t}_{n}^{\star}\|_2 \leq \big\|{\nabla}f_{2}\big({\mathbf{t}}^{(t)}_{n}\big) / \eta^{(t)}_{n}\big\|_2 \label{eq:circle_constraint} \\
			\ & \|\mathbf{t}_{n} - \mathbf{t}_{n}^{k}\|_2 \leq \big\|{\nabla}f_{3,k}\big({\mathbf{t}}^{(t)}_{n}\big) / \upsilon^{(t)}_{k,n} \big\|_2, \ \ \forall k\\
			\ & \eqref{cons:ap_size}. \notag
		\end{align}
	\end{subequations}
	
	\item \textbf{Case 3: Active Points Within Radius \(\big(D+ \big\|{\nabla}f_{2}\big({\mathbf{t}}^{(t)}_{n}\big) / \eta^{(t)}_{n}\big\|\big)\)} 
	 
	If there are active points within a radius \(\big(D + \big\|{\nabla}f_{2}\big({\mathbf{t}}^{(t)}_{n}\big) / \eta^{(t)}_{n}\big\|\big)\) centered at \(\mathbf{t}_{n}^{\star}\), the geometric boundary method proposed in Algorithm~\ref{alg:geometric_boundary} is applied. The feasible solutions are then filtered to identify the point closest to \(\mathbf{t}_{n}^{\star}\) that satisfies all constraints, including \eqref{cons:P17_min_rate}, \eqref{cons:ap_size} and \eqref{cons:max_ap_lbs}.
\end{enumerate}
\begin{algorithm}[tb]
	\SetAlgoLined
	\SetKwInOut{Input}{Input}\SetKwInOut{Output}{Output}
	
	\caption{FP-BCD Algorithm for Beamforming and Antenna Design}
	\label{alg:fp_ao}
	
	\Input{CSI(AoD/AoAs, PRM, and noise power)}
	\Output{$\tilde{\mathbf{t}}, \tilde{\mathbf{r}}, \mathbf{W}_{\mathrm{T}}, \mathbf{w}_{\mathrm{L}},  \mathbf{v}_{\mathrm{A}}$}
	
	\BlankLine
	Randomly initialize $\tilde{\mathbf{t}}, \tilde{\mathbf{r}}, \mathbf{W}_{\mathrm{T}}, \mathbf{w}_{\mathrm{L}},  \mathbf{v}_{\mathrm{A}}$\; 
	\Repeat{The objective value \eqref{eq:opt_1} converges}{
		Calculate $\tilde{\mathbf{y}}$ and $\mathbf{z}$ according to \eqref{eq:opt_y} and \eqref{eq:opt_z}\;
		Update $\mathbf{w}_{\mathrm{L}}$ according to \eqref{eq:opt_lbs_beam}\;
		Update $\mathbf{W}_{\mathrm{T}}$ by solving $\mathbf{P10}$\;
		
		\Repeat{The objective value \eqref{eq:opt_11} converges}{
			Calculate $\bm{\mu}$ according to \eqref{eq:opt_mu}\;
			Update $\mathbf{v}_{\mathrm{A}}$ according to \eqref{eq:opt_v_A}\;
			\Repeat{The objective value \eqref{eq:opt_16} converges}{
				Calculate $\kappa$ according to \eqref{eq:opt_kappa}\;
				Update ${\mathbf{r}}_{m}$ based on the \textbf{Algorithm \ref{alg:geometric_boundary}};
			}
		}
		\Repeat{The objective value \eqref{obj:prob16} converges}{
			Update ${\mathbf{t}}_{n}$ by solving the following cases:
			\begin{itemize}
				\item Case 1: ${\mathbf{t}}_{n} \gets \mathbf{t}_{n}^{\star}$;
				\item Case 2: Solve \textbf{Problem \eqref{eq:prob_t}} to obtain ${\mathbf{t}}_{n}$;
				\item Case 3: Update ${\mathbf{t}}_{n}$ via \textbf{Algorithm \ref{alg:geometric_boundary}};
			\end{itemize}
		}
		
		
	}
\end{algorithm}
\vspace{-5mm}
\subsection{Proposed Algorithm and Complexity Analyses}\label{subsec:complexity}
The proposed FP-BCD algorithm is constructed based on the derivations discussed above and is presented in Algorithm \ref{alg:fp_ao}. 
In the iterations, the beamforming vectors and APVs are updated sequentially. Before each update of the beamforming vector, the auxiliary variables, $\left\{\tilde{y}_{k}, z_{k}\right\}$, are updated once to ensure the equivalence between the original objective function of $\mathbf{P1}$ and the transformed objective function of $\mathbf{P5}$. When updating the APVs, the geometric boundary method is performed to to satisfy the minimum distance constraint, as included in Step 9 and Step 16.

Then, we provide the convergence and analysis for the algorithm of each subproblem. The semi-closed-form solutions for $\mathbf{w}_{\mathrm{L}}$ and $\mathbf{W}_{\mathrm{T}}$ ensure monotonic objective improvement during iterative optimization. Furthermore, the geometric boundary method employed for optimizing $\tilde{\mathbf{t}}$ and $\tilde{\mathbf{r}}$ guarantees a non-decreasing sequence of objective values during iterative updates due to the strict surrogate function properties. Thus, combined with the fact that the optimization of the combiner $\mathbf{v}_{\mathrm{A}}$ is the optimal closed-form solution, we can obtain
\begin{equation}\label{eq:seq}
	\begin{aligned}\
		& R\left(\mathbf{w}_{\mathrm{L}}^{(i_d)}, \mathbf{W}_{\mathrm{T}}^{(i_d)}, \mathbf{v}_{\mathrm{A}}^{(i_d)}, \tilde{\mathbf{r}}^{(i_d)}, \tilde{\mathbf{t}}^{(i_d)}\right) \\
		&\stackrel{}{\leq} R\left(\mathbf{w}_{\mathrm{L}}^{(i_d+1)}, \mathbf{W}_{\mathrm{T}}^{(i_d)}, \mathbf{v}_{\mathrm{A}}^{(i_d)}, \tilde{\mathbf{r}}^{(i_d)}, \tilde{\mathbf{t}}^{(i_d)}\right) \\
		&\stackrel{}{\leq} R\left(\mathbf{w}_{\mathrm{L}}^{(i_d+1)}, \mathbf{W}_{\mathrm{T}}^{(i_d+1)}, \mathbf{v}_{\mathrm{A}}^{(i_d)}, \tilde{\mathbf{r}}^{(i_d)}, \tilde{\mathbf{t}}^{(i_d)}\right) \\
		&\stackrel{}{\leq} R\left(\mathbf{w}_{\mathrm{L}}^{(i_d+1)}, \mathbf{W}_{\mathrm{T}}^{(i_d+1)}, \mathbf{v}_{\mathrm{A}}^{(i_d+1)}, \tilde{\mathbf{r}}^{(i_d+1)}, \tilde{\mathbf{t}}^{(i_d)}\right) \\
		&\stackrel{}{\leq} R\left(\mathbf{w}_{\mathrm{L}}^{(i_d+1)}, \mathbf{W}_{\mathrm{T}}^{(i_d+1)}, \mathbf{v}_{\mathrm{A}}^{(i_d+1)}, \tilde{\mathbf{r}}^{(i_d+1)}, \tilde{\mathbf{t}}^{(i_d+1)}\right)
	\end{aligned}
\end{equation}
where $R\left(\mathbf{w}_{\mathrm{L}}, \mathbf{W}_{\mathrm{T}}, \mathbf{v}_{\mathrm{A}}, \tilde{\mathbf{r}}, \tilde{\mathbf{t}}\right)$ is the objective function of $\mathbf{P1}$. Hence, the sequence \eqref{eq:seq} is monotonically increasing. Furthermore, $R\left(\mathbf{w}_{\mathrm{L}}, \mathbf{W}_{\mathrm{T}}, \mathbf{v}_{\mathrm{A}}, \tilde{\mathbf{r}}, \tilde{\mathbf{t}}\right)$ is upper-bounded due to $\eqref{cons:ap_size}- \eqref{cons:min_user_rate}$, which leads to the convergence of the proposed algorithm.

Next, the computational complexity of the proposed algorithm is presented. For the optimization of $\mathbf{w}_{\mathrm{L}}$, the complexity is given by $\mathcal{O}\left(\left(N_{\mathrm{L}}+K+1\right)^2 + N_{\mathrm{L}}\right)$ according to \cite{syf_ris_aj_1}. As for the design of $\mathbf{w}_{\mathrm{T}}$, it is divided into two steps. The first step includes determining beamforming direction, with a complexity of $\mathcal{O}\left(KN^{2}_{\mathrm{T}} + N^{3}_{\mathrm{T}}\right)$; the second step is conducted for power allocation, with a complexity of $\mathcal{O}\left(K^{3.5}\right)$. Therefore, the complexity of optimizing $\mathbf{w}_{\mathrm{T}}$ can be obtained as $\mathcal{O}\left(KN^{2}_{\mathrm{T}} + N^{3}_{\mathrm{T}}+K^{3.5}\right)$. Then, as for the optimization of $\mathbf{v}_{\mathrm{A}}$ and $\tilde{\mathbf{r}}$ by using AO-based optimization framework, their total complexity is computed as 
\begin{equation}
	\begin{aligned}
		 \mathcal{O}( & \log_{2}(\varepsilon^{-1})(N^{3}_{\mathrm{A}} + T_{1}N_{\mathrm{A}}(\mathrm{max}\left\{N_{\mathrm{T}}L^{2}_{\mathrm{TA}}, QN_{\mathrm{J}}L^{2}_{\mathrm{JA}}, N_{\mathrm{L}}L^{2}_{\mathrm{LA}}\right\} \\ & + \left(N_{\mathrm{A}}-1\right)^{2} + \log_{2}(N^{2}_{\mathrm{A}}-1)))) \notag
	\end{aligned}
\end{equation}
where $\varepsilon$ is the accuracy and $T_{1}$ denotes the iteration number of CCD algorithm. Finally, the complexity for optimizing $\tilde{\mathbf{t}}$ is 
$$
	\mathcal{O}\left(T_{2}N_{\mathrm{L}}\left(N_{\mathrm{A}}L^{2}_{\mathrm{LA}} \! + \!KL^{2}_{\mathrm{L},k} \! + \! \left(N_{\mathrm{L}} \!- \!1\right)^{2}  \!+  \!\log_{2}\left(N^{2}_{\mathrm{L}} \!- \!1\right)\right)\right)
$$
where $T_{2}$ denotes the iteration number. Thus, the total complexity of proposed optimization framework is obtained as 
\begin{equation}
	\begin{aligned}
		&\mathcal{O}_{\mathrm{tot}} \\  
		& = \mathcal{O}\Big(\mathrm{max}\Big\{\left(N_{\mathrm{L}}+K+1\right)^2 + N_{\mathrm{L}}, KN^{2}_{\mathrm{T}} + N^{3}_{\mathrm{T}}+K^{3.5}, \\
		& \log_{2}(\varepsilon^{-1})\big(N^{3}_{\mathrm{A}} + T_{1}N_{\mathrm{A}}\big(\mathrm{max}\left\{N_{\mathrm{T}}L^{2}_{\mathrm{TA}}, QN_{\mathrm{J}}L^{2}_{\mathrm{JA}}, N_{\mathrm{L}}L^{2}_{\mathrm{LA}}\right\} \\
		&+ \left(N_{\mathrm{A}}-1\right)^{2} + \log_{2}(N^{2}_{\mathrm{A}}-1)\big)\big), T_{2}N_{\mathrm{L}}\big(N_{\mathrm{A}}L^{2}_{\mathrm{LA}} + KL^{2}_{\mathrm{L},k}\\
		&  + \left(N_{\mathrm{L}}-1\right)^{2} + \log_{2}\left(N^{2}_{\mathrm{L}}-1\right)\big) \Big\}\Big)
	\end{aligned} \notag
\end{equation}

\section{Numerical Results}\label{sec:num_results}
In this section, we present numerical results to demonstrate the performance of the proposed algorithm
.
\vspace{-3mm}
\subsection{Simulation Settings}
We consider a scenario with a jammer equipped with $N_{\mathrm{J}}=16$ FPAs. The TBS is equipped with $N_{\mathrm{G}}=16$ FPAs. and serves $K=4$ single-antenna terrestrial users simultaneously. The LBS is equipped with $N_{\mathrm{L}}=16$ fluid antennas and serves an aerial user equipped with $N_{\mathrm{A}}=4$ fluid antennas. The transmit and receive regions are set as square areas with size $A \times A$, with $A=3\lambda$. In addition, the information rate target threshold for terrestrial user $k$ is set as $R_{k,\mathrm{min}}=1$ bps/Hz. The maximum transmit power of the LBS is set to $P_{\mathrm{L,max}} = 10 \ \text{dBm}$. Additionally, the total jamming power is set to $p_{\mathrm{J}} = 30 \ \text{dBm}$, and the CSI uncertainty is defined as $\Delta = \theta_U-\theta_L=\phi_U-\phi_L = 4^{\circ}$. The carrier frequency is set to 3 GHz (with a corresponding wavelength of $\lambda=$ 0.1 m), and the minimum distance $D$ between adjacent antennas is half the wavelength. The TBS is located at the origin with a height of 10 meters, while the LBS is positioned at (0 m, 100 m, 10 m). The aerial user and the jammer are located at (10 m, 60 m, 100 m) and (40 m, 80 m, 10 m), respectively. The terrestrial users are randomly distributed within a circle of radius 10 meters, centered at (40 m, 30 m, 1.5 m).

Assume that the channel from the jammer to the terrestrial users is blocked. The channels involving TBS to terrestrial user $k$, TBS to aerial user, LBS to terrestrial user $k$, LBS to aerial user, and jammer to aerial user are modeled as geometric channel models. For simplicity, we assume that all users have the same number of transmit and receive paths, denoted as $L$. The path response matrix is assumed to be diagonal with $\bm{\Sigma}\left[1,1\right] \sim \mathcal{CN}\left(0,10^{PL/10}\right)$ and $\bm{\Sigma}\left[l,l\right] \sim \mathcal{CN}\left(0,10^{PL/10}/(L-1)\right)$ for $l=2,3,\dots,L$, where $PL=-40-2.8\log_{10}(d)(\text{dB})$ and $d$ is the link distance in meters. Furthermore, the noise variance of the received signals is assumed to be identical across all users, set to $\sigma^{2}_{\mathrm{T},1}=\sigma^{2}_{\mathrm{T},2}=\cdots=\sigma^{2}_{\mathrm{T},K}=\sigma^{2}_{\mathrm{A}}=-80 \ \text{dBm}$. 
\begin{figure}[t]\vspace{-0mm}
	\begin{center}		\centerline{\includegraphics[width=0.3\textwidth]{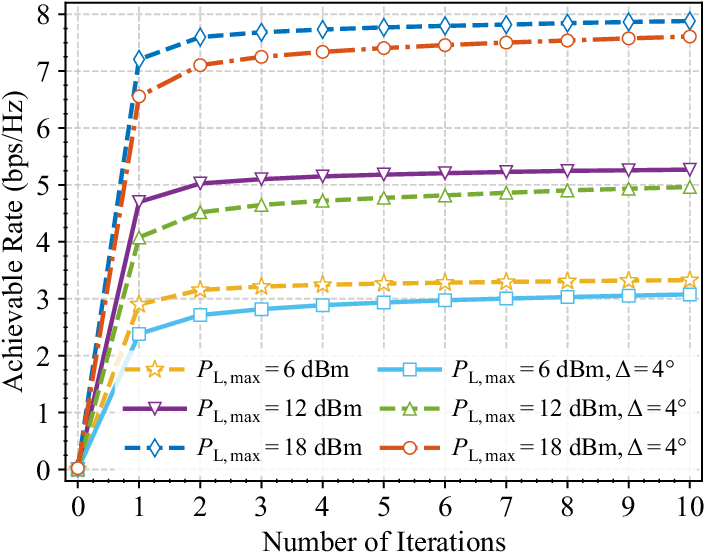}}  \vspace{-0mm}
		\captionsetup{font=footnotesize, name={Fig.}, labelsep=period}  
		\caption[t]{\raggedright Achievable rate versus number of iterations.}
		\label{res:conve}
	\end{center}
	\vspace{-5mm}
\end{figure}

To serve as baselines, other schemes are listed below.
\begin{itemize}
	\item \textbf{Baseline 1} (FPA): The LBS and aerial user are equipped with FPA-based uniform planar arrays (UPAs) with $N_{\mathrm{L}}$ and $N_{\mathrm{A}}$ antennas, respectively, spaced by $\lambda/2$.
	\item \textbf{Baseline 2} (Transmit FAS): The LBS is equipped with $N_{\mathrm{L}}$ fluid antennas, while the aerial user is equipped with an FPA-based UPA same as the FPA scheme.
	\item \textbf{Baseline 3} (Receive FAS): The aerial user is equipped with $N_{\mathrm{A}}$ fluid antennas, while the LBS is equipped with an FPA-based UPA same as the FPA scheme.
	\item \textbf{Baseline 4} (FAS-AO): The traditional method in \cite{TWC_MWY} is adopted, which optimizes antenna positions with CVX alternatively and greedily.
	\item \textbf{Baseline 5} (FAS-PGD): The method employs projected gradient descent (PGD) with Armijo line search \cite{TWC_XZY} to jointly optimize all antenna positions, ensuring both convergence and constraint satisfaction.
	\item \textbf{Baseline 6} (FAS-AS): The method employs an antenna selection (AS) approach that discretizes the predefined space into candidate positions spaced at $\lambda/2$ intervals, then alternately optimizes to select the optimal discrete position for each antenna.
\end{itemize}
\vspace{-2mm}
\subsection{Convergence Analysis}
First, the proposed algorithm is evaluated for its convergence properties. We consider varying maximum transmit powers and CSI uncertainty as examples, Fig. \ref{res:conve} illustrates the achievable rate plotted against the number of iterations. The performance of the proposed FP-BCD algorithm improves with increasing iteration number due to its non-decreasing property, as elaborated in \ref{subsec:complexity}. In the 10th iteration, under a CSI uncertainty $\Delta$ of $4^{\circ}$ and a transmit power of 12 dBm, the proposed algorithm exhibits only a 5.78\% performance loss compared to perfect CSI, demonstrating the strong robustness of FP-BCD. When comparing the iterations required for convergence, it is evident that beamforming with higher maximum transmit power does not necessitate more iterations, whereas the presence of CSI uncertainty $\Delta$ demands additional iterations. This requirement arises from the increased jamming discretization samples in scenarios with CSI uncertainty. Unless otherwise specified, the number of iterations for FP-BCD in subsequent tests is set to 10.
\subsection{Performance Analysis}
\begin{figure}[!t]
	\hspace{-1mm}
	\begin{subfigure}[b]{0.24\textwidth}
		\includegraphics[width=\textwidth]{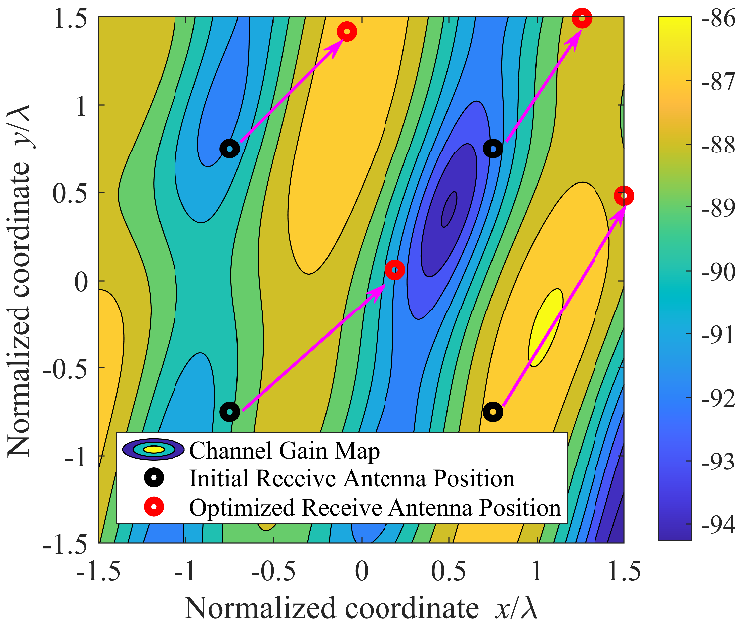}
		\captionsetup{font=footnotesize, name={Fig.}, labelsep=period}  
		\caption{}
		\label{fig:sub_a}
	\end{subfigure}
	\hspace{-2.5mm}
	\begin{subfigure}[b]{0.24\textwidth}
		\includegraphics[width=\textwidth]{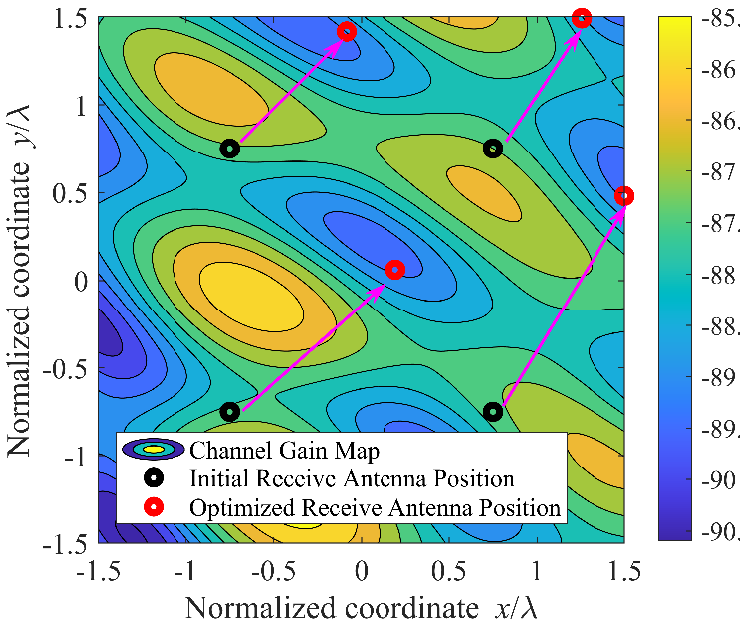}
		\captionsetup{font=footnotesize, name={Fig.}, labelsep=period}  
		\caption{}
		\label{fig:sub_b}
	\end{subfigure}
	\captionsetup{font=footnotesize, name={Fig.}, labelsep=period}  
	\caption[t]{\raggedright Channel power gains (dB) (a) between the transmit fluid antenna of the LBS and the receive fluid antenna of the aerial user, (b) between the FPA of the jammer and the receive fluid antenna of the aerial user.}
	\label{res:channel_gain}
\end{figure}

To validate the impact of APV optimization on improving channel conditions and anti-jamming performance, Fig. \ref{res:channel_gain} presents the realized channel power gain (in dB) for $\mathbf{H}_{\mathrm{LA}}$ and $\mathbf{H}_{\mathrm{JA}}$ within the movement region of the receiving antennas. APV optimization should preferentially select positions with better channel conditions for legitimate users while choosing positions with worse channel conditions for malicious jammers. For numerical verification of our proposed algorithm, we define the channel power gain as $\Vert \mathbf{H}_{\mathrm{LA}} \Vert^{2}$ and $\Vert \mathbf{H}_{\mathrm{JA}} \Vert^{2}$. By employing the proposed algorithm, the channel power gain between the LBS and the aerial user increases from -83.1 dB to -81.7 dB, whereas the channel power gain between the jammer and the aerial user decreases from -81.1 dB to -82.9 dB. This indicates that by relocating the antennas, the channel gain of the jammer can be effectively suppressed while enhancing that of the legitimate user.
\begin{figure*}[t]
	\centering
	\begin{subfigure}[b]{0.3\textwidth}
		\includegraphics[width=\textwidth]{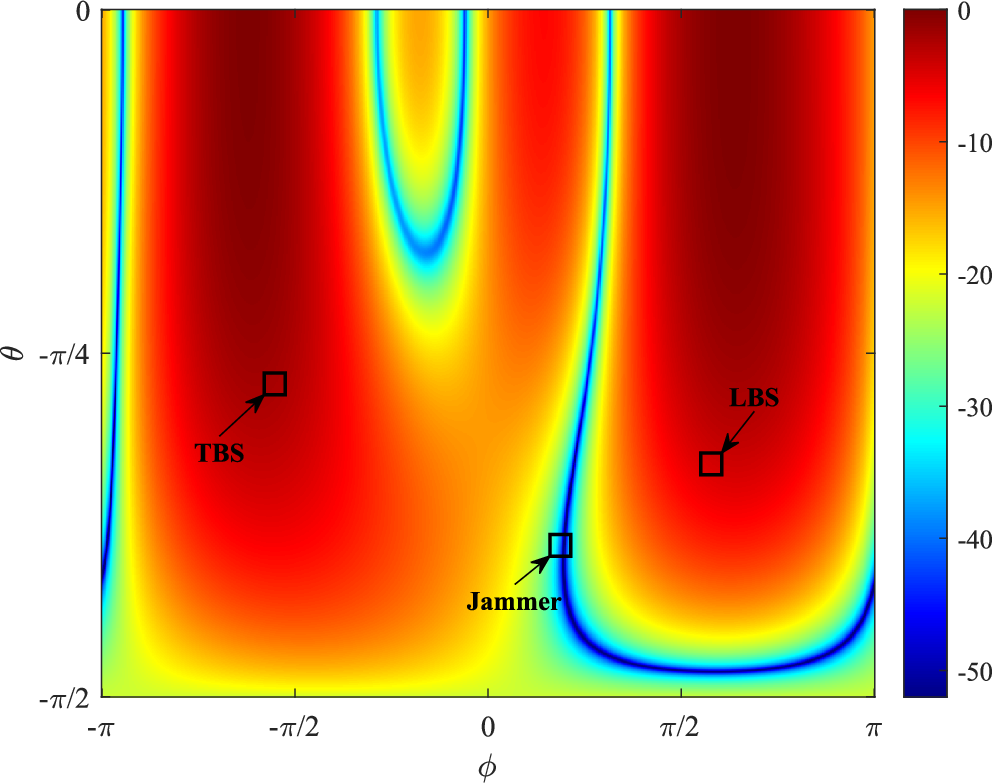}
		\captionsetup{font=footnotesize, name={Fig.}, labelsep=space}  
		\caption{2D beampattern of the FPA.}
		\label{fig:beam_sub_a}
	\end{subfigure}
	\hspace{8mm}
	\begin{subfigure}[b]{0.3\textwidth}
		\includegraphics[width=\textwidth]{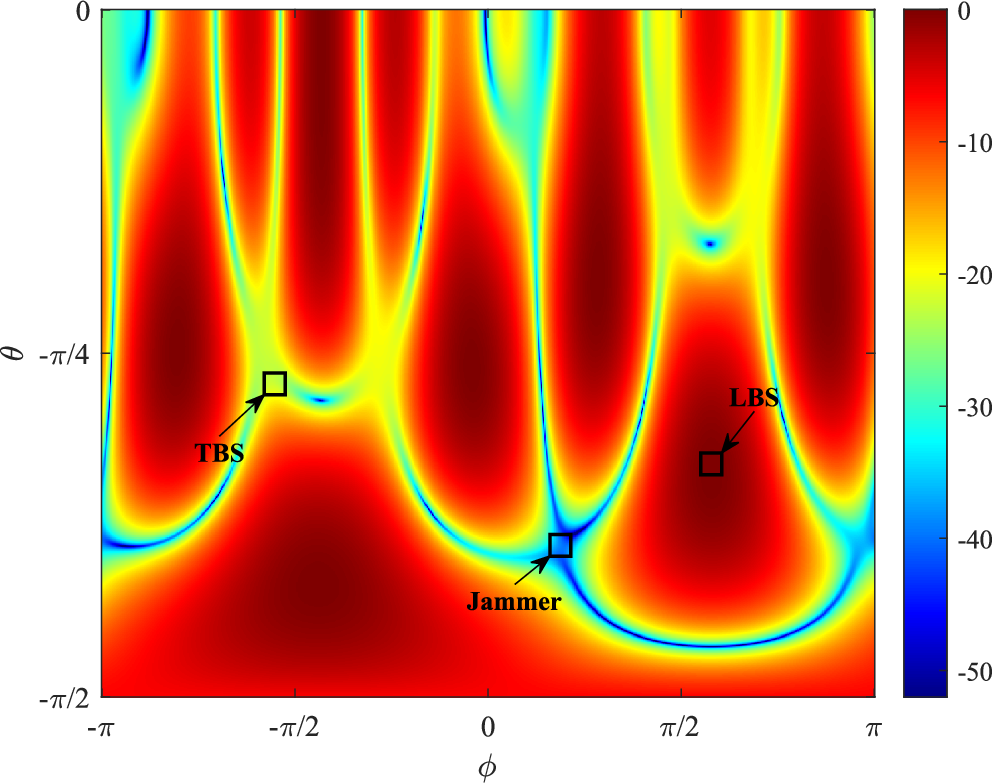}
		\captionsetup{font=footnotesize, name={Fig.}, labelsep=space}  
		\caption{2D beampattern of the FAS.}
		\label{fig:beam_sub_b}
	\end{subfigure}
	\captionsetup{font=footnotesize, name={Fig.}, labelsep=period}  
	\caption[t]{\raggedright Receive beampattern with different architectures (colorbar on the right is unified for 2 subfigures, unit: dB).}
	\label{res:beam_pattern}
	\vspace{-4mm}
\end{figure*}

\begin{figure*}[t]
	\centering
	\begin{subfigure}[b]{0.3\textwidth}
		\includegraphics[width=\textwidth]{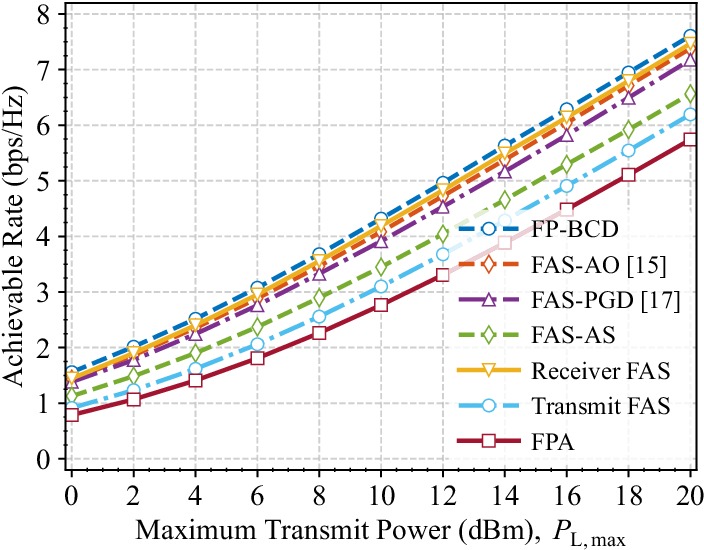}
		\captionsetup{font=footnotesize, name={Fig.}, labelsep=space}  
		\caption{}
		\label{fig:a}
	\end{subfigure}
	\hspace{4mm}
	\begin{subfigure}[b]{0.3\textwidth}
		\includegraphics[width=\textwidth]{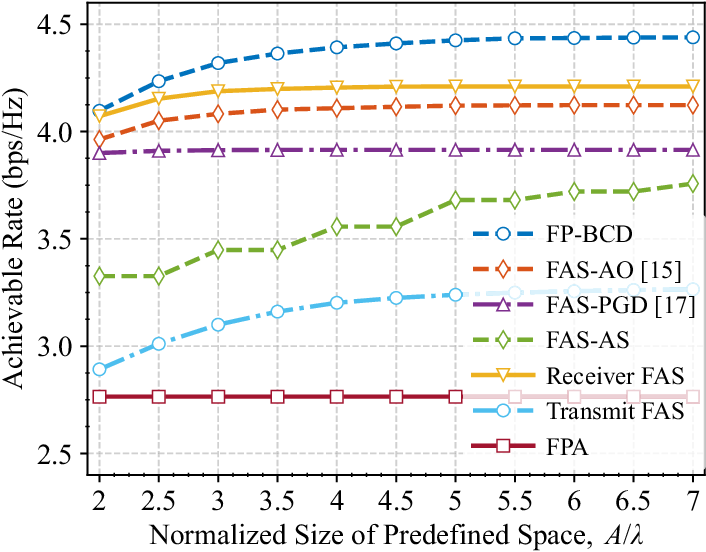}
		\captionsetup{font=footnotesize, name={Fig.}, labelsep=space}  
		\caption{}
		\label{fig:b}
	\end{subfigure}
	\hspace{4mm}
	\begin{subfigure}[b]{0.3\textwidth}
		\includegraphics[width=\textwidth]{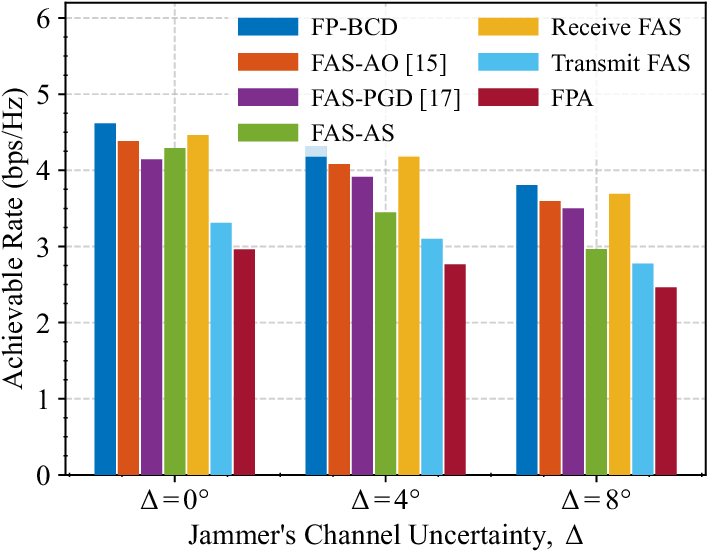}
		\captionsetup{font=footnotesize, name={Fig.}, labelsep=space}  
		\caption{}
		\label{fig:c}
	\end{subfigure}
	\captionsetup{font=footnotesize, name={Fig.}, labelsep=period}  
	\caption[t]{\raggedright Achievable rate with different system parameters, (a) maximum transmit power, $P_{\mathrm{L,max}}$, (b) normalized size of predefined space, $A/\lambda$, (c) jammer's channel uncertainty, $\Delta$}
	\label{res:rate}
	\vspace{-4mm}
\end{figure*}

Fig. \ref{res:beam_pattern} shows the normalized receive beampattern of different architectures and evaluates the quality of the beam by comparing their mainlobes and nulls. The proposed algorithm enables the aerial user's receiver to accurately generate nulls toward the jammer region while simultaneously aligning the main lobe with the desired target, even under angle uncertainty. Furthermore, it is noteworthy that the FAS receiver achieves an SINR of approximately 0 dB in the LBS direction and -22 dB in the TBS direction, while the FPA receiver yields corresponding SINRs of -4.5 dB and -2 dB, respectively. These results indicate that the FAS receiver can effectively suppress undesired signals while enhancing the desired ones. Additionally, both the FPA and FAS receivers are capable of forming deep nulls in the direction of the jammer, maintaining received SINRs of approximately -40 dB. These observations confirm that the proposed FAS transceiver can fully exploit additional spatial DoFs to enhance anti-jamming performance.

Subsequently, we evaluate the performance of the proposed methods under varying maximum transmit power. Fig. \ref{fig:a} depict the achievable rate as function of the maximum transmit power, $P_{\mathrm{L,max}}$. Experimental results demonstrate that the proposed FP-BCD algorithm exhibits superior performance compared to state-of-the-art algorithms. The scheme with FAS deployed at both transmitter and receiver achieves a rate of 4.32 bps/Hz, whereas the FPA scheme only reaches 2.76 bps/Hz when $P_{\mathrm{L,max}}$ = 10 dBm. This indicates that APV optimization provides a 56\% performance gain. Notably, the receiver-side FAS deployment scheme performs comparably to the dual-end FAS deployment scheme, and both significantly surpass both the transmitter-side FAS deployment scheme and the FPA scheme. This enhancement primarily stems from the fact that in scenarios with strong-power jamming, the achievable rate predominantly depends on the spatial-domain suppression capability against jamming signals.

Next, we conduct a comprehensive performance evaluation of the proposed FP-BCD algorithm across varying normalized sizes. As illustrated in Fig. \ref{fig:b}, the proposed FP-BCD algorithm consistently outperforms other algorithms across various normalized sizes. The performance of all FAS schemes improves with increasing normalized region size, attributed to the fact that expanding the size enables more comprehensive exploitation of spatial DoF, thereby achieving more effective jamming suppression and legitimate channel gain enhancement through dynamic APV optimization. It is particularly noteworthy that due to the finite multipath components in practical propagation environments, the channel gain displays characteristic periodicity \cite{TWC_MA_Model}. When the region size exceeds a critical threshold, FAS can sufficiently capture the spatial variation characteristics of the channel, at which point the system achievable rate converges to a stable value without further improvement with increasing size.

Fig. \ref{fig:c} shows the achievable rate versus the jamming channel uncertainty $\Delta$. It can be seen that the achievable rate decreases with $\Delta$, and the scheme with FAS deployed at both transmitter and receiver consistently maintains performance superiority. Particularly noteworthy is that the receiver-side FAS deployment scheme demonstrates higher sensitivity to $\Delta$ variation compared to other schemes, whereas the FPA scheme exhibits the strongest robustness. This phenomenon can be attributed to the fact that the receiver-side FAS scheme requires precise matching of the AoAs of jamming signals for APV design, while the FPA scheme, due to its limited spatial DoF, possesses relatively stronger adaptability to channel uncertainty.

\section{Conclusion}
This work has demonstrated significant performance gains achieved by FAS deployment in low-altitude anti-jamming communications. We propose a dual-end FAS architecture, with strategic antenna repositioning at both transmitter and receiver, which enhances spatial interference suppression while boosting legitimate signal strength. We establish FP-BCD algorithm that coordinates beamforming, combining, and antenna positioning through block coordinate descent, while the geometric boundary method efficiently resolves the antenna positioning problem. Experimental validation confirms higher achievable rates compared to FPA systems under equivalent power constraints, alongside robust operation across diverse jamming uncertainties. Future work will extend this approach to multi-user FAS networks.

\appendix
\begin{figure*}[hb]
	\centering
	\vspace{-4mm}
	\hrulefill
	\begin{subequations}\label{eq:hess}
		\begin{align}
			\frac{\partial^{2}f_{1}}{\partial^{2}x^{\mathrm{r}}_{m}} &= \frac{8\pi^{2}}{\lambda^{2}}\sum\limits_{\hat{k}=1}^{Q'}\bar{\mu}_{\hat{k}}\sum_{l=1}^{L^{\mathrm{r}}_{\mathrm{A},\hat{k}}}{\bar{\kappa}_{\hat{k}}}\left|v_{\mathrm{A},m}\right|^{2}\mathrm{Re}\left\{\tilde{\bm{\Sigma}}^{*}_{\mathrm{A},\hat{k},l}{\mathbf{\dot{f}}}^{*}_{\mathrm{A},\hat{k},m}\dot{f}_{\mathrm{A},\hat{k},m,l}\right\} - \mathrm{Re}\left\{\bar{f}^{*}_{\mathrm{A},\hat{k},m,l}\ddot{f}_{\mathrm{A},\hat{k},m,l}\right\} \\
			\frac{\partial^{2}f_{1}}{\partial x^{\mathrm{r}}_{m}\partial y^{\mathrm{r}}_{m}} = \frac{\partial^{2}f_{1}}{\partial y^{\mathrm{r}}_{m}\partial x^{\mathrm{r}}_{m}} &= \frac{8\pi^{2}}{\lambda^{2}}\sum\limits_{\hat{k}=1}^{Q'}\bar{\mu}_{\hat{k}}\sum_{l=1}^{L^{\mathrm{r}}_{\mathrm{A},\hat{k}}}{\bar{\kappa}_{\hat{k}}}\left|v_{\mathrm{A},m}\right|^{2}\mathrm{Re}\left\{\tilde{\bm{\Sigma}}^{*}_{\mathrm{A},k,l}{\mathbf{f'}}^{*}_{\mathrm{A},\hat{k},m}\dot{f}_{\mathrm{A},\hat{k},m,l}\right\}  - \mathrm{Re}\left\{\bar{f}^{*}_{\mathrm{A},\hat{k},m,l}\dot{f}'_{\mathrm{A},\hat{k},m,l}\right\} \\
			\frac{\partial^{2}f_{1}}{\partial^{2}y^{\mathrm{r}}_{m}} &= \frac{8\pi^{2}}{\lambda^{2}}\sum\limits_{\hat{k}=1}^{Q'}\bar{\mu}_{\hat{k}}\sum_{l=1}^{L^{\mathrm{r}}_{\mathrm{A},\hat{k}}}{\bar{\kappa}_{\hat{k}}}\left|v_{\mathrm{A},m}\right|^{2}\mathrm{Re}\left\{\tilde{\bm{\Sigma}}^{*}_{\mathrm{A},\hat{k},l}{\mathbf{f'}}^{*}_{\mathrm{A},\hat{k},m}{f}'_{\mathrm{A},\hat{k},m,l}\right\}  - \mathrm{Re}\left\{\bar{f}^{*}_{\mathrm{A},\hat{k},m,l}{f}''_{\mathrm{A},\hat{k},m,l}\right\}
		\end{align}
	\end{subequations}
\end{figure*}
\subsection{Derivations of $\nabla f_{1}\left(\mathbf{r}_{m}\right)$ and $\nabla^{2} f_{1}\left(\mathbf{r}_{m}\right)$}\label{app:A}
Plugging the the channel model into $f_{1}$, we can obtain
\begin{equation}
	\begin{aligned}
		& f_{1}  \!=\!  \kappa \sum_{i=1}^{N_{\mathrm{A}}}\sum_{j=1}^{N_{\mathrm{A}}}v^{*}_{\mathrm{A},i}v_{\mathrm{A},j} \mathbf{f}^{\mathrm{H}}_{\mathrm{GA}}\left(\mathbf{r}_{i}\right)\tilde{\bm{\Sigma}}_{\mathrm{GA}}\mathbf{f}_{\mathrm{GA}}\left(\mathbf{r}_{j}\right) \\
		& \!+ \!\kappa \sigma^{2}_{\mathrm{A}} - \sum_{i=1}^{N_{\mathrm{A}}}\sum_{j=1}^{N_{\mathrm{A}}}v^{*}_{\mathrm{A},i}v_{\mathrm{A},j}\mathbf{f}^{\mathrm{H}}_{\mathrm{LA}}\left(\mathbf{r}_{i}\right)\tilde{\bm{\Sigma}}_{\mathrm{LA}}\mathbf{f}_{\mathrm{LA}}\left(\mathbf{r}_{j}\right)\\
		&\! +\! \kappa \hat{p}_{\mathrm{J}}\! \sum_{i=1}^{N_{\mathrm{A}}}\sum_{j=1}^{N_{\mathrm{A}}}\!v^{*}_{\mathrm{A},i}v_{\mathrm{A},j} \bigg( \sum_{p=1}^{Q} \mu_{p} \mathbf{f}_{\mathrm{JA}}^{(p),\mathrm{H}}\big(\mathbf{r}_{i}\big) \tilde{\bm{\Sigma}}_{\mathrm{JA}} \mathbf{f}^{(p)}_{\mathrm{JA}}\big(\mathbf{r}_{j}\big) \!\bigg) \notag
	\end{aligned}
\end{equation}
where $\bar{\bm{\Sigma}}_{\mathrm{GA}} \!=\! {\bm{\Sigma}}_{\mathrm{GA}}{\mathbf{G}_{\mathrm{GA}}}$, $\tilde{\bm{\Sigma}}_{\mathrm{GA}} \!=\! \bar{\bm{\Sigma}}_{\mathrm{GA}}\mathbf{W}_{\mathrm{G}}\mathbf{W}^{\mathrm{H}}_{\mathrm{G}}\bar{\bm{\Sigma}}^{\mathrm{H}}_{\mathrm{GA}}$, $\bar{\bm{\Sigma}}_{\mathrm{JA}} \!=\! {\bm{\Sigma}}_{\mathrm{JA}}{\mathbf{G}_{\mathrm{JA}}}$, $\tilde{\bm{\Sigma}}_{\mathrm{JA}}=\bar{\bm{\Sigma}}_{\mathrm{JA}}\bar{\bm{\Sigma}}^{\mathrm{H}}_{\mathrm{JA}}$, $\bar{\bm{\Sigma}}_{\mathrm{LA}} = {\bm{\Sigma}}_{\mathrm{LA}}{\mathbf{G}_{\mathrm{LA}}}$, $\tilde{\bm{\Sigma}}_{\mathrm{LA}} =\bar{\bm{\Sigma}}_{\mathrm{LA}}\mathbf{w}_{\mathrm{L}}\mathbf{w}^{\mathrm{H}}_{\mathrm{L}}\bar{\bm{\Sigma}}^{\mathrm{H}}_{\mathrm{LA}}$.

The gradient of $f_{1}$ with respect to $\mathbf{r}_{m}$, incorporating TBS ($\hat{k}=1$), LBS ($\hat{k}=2$), and jammer terms ($\hat{k}=3,\ldots,Q'$, with $Q' = Q+ 2$), is given by
\begin{equation}
	\!\!\!\!\nabla f_{1}(\mathbf{r}_{m}) \!=\! -\frac{4\pi}{\lambda}\!\! \begin{bmatrix}
		\sum\limits_{\hat{k}=1}^{Q'} \!\bar{\mu}_{\hat{k}} \sum\limits_{l=1}^{L^{\mathrm{r}}_{\mathrm{A},\hat{k}}}\! \mathrm{Im} \left\{ \bar{f}^{*}_{\mathrm{A},\hat{k},m,l} \dot{f}_{\mathrm{A},\hat{k},m,l} \right\} \\
		\sum\limits_{\hat{k}=1}^{Q'} \!\bar{\mu}_{\hat{k}} \sum\limits_{l=1}^{L^{\mathrm{r}}_{\mathrm{A},\hat{k}}}\! \mathrm{Im} \left\{ \bar{f}^{*}_{\mathrm{A},\hat{k},m,l} {f}'_{\mathrm{A},\hat{k},m,l} \right\}
	\end{bmatrix}
\end{equation}
where $L^{\mathrm{r}}_{\mathrm{A},1} = L^{\mathrm{r}}_{\mathrm{GA}}$, $L^{\mathrm{r}}_{\mathrm{A},2} = L^{\mathrm{r}}_{\mathrm{LA}}$, $L^{\mathrm{r}}_{\mathrm{A},\hat{k}} = L^{\mathrm{r}}_{\mathrm{JA}} $, for all $\hat{k} > 2$, and
\begin{align}
	\dot{f}_{\mathrm{A},\hat{k},m,l} &= \cos\theta^{\mathrm{r}}_{\mathrm{A},\hat{k},l} \cos\phi^{\mathrm{r}}_{\mathrm{A},\hat{k},l} e^{j \frac{2\pi}{\lambda} \mathbf{r}^{\mathrm{T}}_{m} \mathbf{n}^{\mathrm{r}}_{\mathrm{A},\hat{k},l}} \\
	{f}'_{\mathrm{A},\hat{k},m,l} &= \cos\theta^{\mathrm{r}}_{\mathrm{A},\hat{k},l} \sin\phi^{\mathrm{r}}_{\mathrm{A},\hat{k},l} e^{j \frac{2\pi}{\lambda} \mathbf{r}^{\mathrm{T}}_{m} \mathbf{n}^{\mathrm{r}}_{\mathrm{A},\hat{k},l}}
\end{align}
with AoAs and normalized wave vectors defined as
\begin{itemize}
	\item $\hat{k}=1$: $\theta^{\mathrm{r}}_{\mathrm{A},1,l} = \theta^{\mathrm{r}}_{\mathrm{GA},l}$,  $\phi^{\mathrm{r}}_{\mathrm{A},1,l} = \phi^{\mathrm{r}}_{\mathrm{GA},l}$, $\mathbf{n}^{\mathrm{r}}_{\mathrm{A},1,l} = \mathbf{n}^{\mathrm{r}}_{\mathrm{GA},l}$.
	\item $\hat{k}=2$: $\theta^{\mathrm{r}}_{\mathrm{A},2,l} = \theta^{\mathrm{r}}_{\mathrm{LA},l}$, $\phi^{\mathrm{r}}_{\mathrm{A},2,l} = \phi^{\mathrm{r}}_{\mathrm{LA},l}$, $\mathbf{n}^{\mathrm{r}}_{\mathrm{A},2,l} = \mathbf{n}^{\mathrm{r}}_{\mathrm{LA},l}$.
	\item $\hat{k}\!>\!\!2$: $\!\theta^{\mathrm{r}}_{\mathrm{A},\hat{k},l} = \theta^{\mathrm{r},(\hat{k}-2)}_{\mathrm{JA},l}$, $\phi^{\mathrm{r}}_{\mathrm{A},\hat{k},l} = \phi^{\mathrm{r},(\hat{k}-2)}_{\mathrm{JA},l}$, $\mathbf{n}^{\mathrm{r}}_{\mathrm{A},\hat{k},l} = \mathbf{n}^{\mathrm{r},(\hat{k}-2)}_{\mathrm{JA},l}$.
\end{itemize}
The coefficients are given by
\begin{align}
	\!\!\bar{\mu}_{\hat{k}} &= \begin{cases} 1, & \hat{k}=1,2 \\ \mu_{\hat{k}-2}, & \hat{k}>2 \end{cases} \\
	\!\!\bar{\mathbf{f}}_{\mathrm{A},\hat{k},m} &= \begin{cases}
		\kappa \sum\limits_{i=1}^{N_{\mathrm{A}}} v_{\mathrm{A},i} v^{*}_{\mathrm{A},m} \tilde{\mathbf{\Sigma}}_{\mathrm{GA}} \mathbf{f}_{\mathrm{GA}}(\mathbf{r}_{i}), & \hat{k}=1 \\
		-\sum\limits_{i=1}^{N_{\mathrm{A}}} v_{\mathrm{A},i} v^{*}_{\mathrm{A},m} \tilde{\mathbf{\Sigma}}_{\mathrm{LA}} \mathbf{f}_{\mathrm{LA}}(\mathbf{r}_{i}), & \hat{k}=2 \\
		\kappa \hat{p}_{\mathrm{J}} \sum\limits_{i=1}^{N_{\mathrm{A}}} v_{\mathrm{A},i} v^{*}_{\mathrm{A},m} \tilde{\mathbf{\Sigma}}_{\mathrm{JA}} \mathbf{f}^{(\hat{k}-2)}_{\mathrm{JA}}(\mathbf{r}_{i}), & \hat{k}>2
	\end{cases}.
\end{align}

Then, the Hessian matrix of $f_{1}$ over $\mathbf{r}_{m}$ is given by
\begin{align}
	\nabla^{2}f_{1}\left({\mathbf{r}_{m}}\right) = \left[ {\begin{array}{*{20}{c}}
			\frac{\partial^{2}f_{1}}{\partial^{2} x^{\mathrm{r}}_{m}}&\frac{\partial^{2}f_{1}}{\partial x^{\mathrm{r}}_{m}\partial y^{\mathrm{r}}_{m}}\\
			\frac{\partial^{2}f_{1}}{\partial y^{\mathrm{r}}_{m}x^{\mathrm{r}}_{n}}&\frac{\partial^{2}f_{1}}{\partial^{2}y^{\mathrm{r}}_{m}}
	\end{array}} \right]
\end{align}
where the expressions of the matrix’s elements are provided in \eqref{eq:hess} at the bottom of the page with
\begin{align}
	\ddot{f}_{\mathrm{A},\hat{k},m,l} &= \cos^{2} \theta^{\mathrm{r}}_{\mathrm{A},\hat{k},l} \cos^{2} \phi^{\mathrm{r}}_{\mathrm{A},\hat{k},l} e^{j \frac{2\pi}{\lambda} \mathbf{r}^{\mathrm{T}}_{m} \mathbf{n}^{\mathrm{r}}_{\mathrm{A},\hat{k},l}} \\
	\dot{f}'_{\mathrm{A},\hat{k},m,l} &= \cos^{2} \theta^{\mathrm{r}}_{\mathrm{A},\hat{k},l} \cos\phi^{\mathrm{r}}_{\mathrm{A},\hat{k},l} \sin\phi^{\mathrm{r}}_{\mathrm{A},\hat{k},l} e^{j \frac{2\pi}{\lambda} \mathbf{r}^{\mathrm{T}}_{m} \mathbf{n}^{\mathrm{r}}_{\mathrm{A},\hat{k},l}} \\
	{f}''_{\mathrm{A},\hat{k},m,l} &= \cos^{2} \theta^{\mathrm{r}}_{\mathrm{A},\hat{k},l} \sin^{2} \phi^{\mathrm{r}}_{\mathrm{A},\hat{k},l} e^{j \frac{2\pi}{\lambda} \mathbf{r}^{\mathrm{T}}_{m} \mathbf{n}^{\mathrm{r}}_{\mathrm{A},\hat{k},l}}.
\end{align}
\begin{equation}
	\tilde{\mathbf{\Sigma}}_{\mathrm{A},\hat{k}} = \begin{cases}
		\tilde{\mathbf{\Sigma}}_{\mathrm{GA}}, & \hat{k}=1 \\
		\tilde{\mathbf{\Sigma}}_{\mathrm{LA}}, & \hat{k}=2 \\
		\tilde{\mathbf{\Sigma}}_{\mathrm{JA}}, & \hat{k}>2
	\end{cases}, \quad
	\bar{\kappa}_{\hat{k}} = \begin{cases}
		\kappa, & \hat{k}=1 \\
		-1, & \hat{k}=2 \\
		\kappa p_{\mathrm{J}}, & \hat{k}>2
	\end{cases}. \notag
\end{equation}
Here, $\tilde{\mathbf{\Sigma}}^{*}_{\mathrm{A},\hat{k},l}$ is the $l$-th row of $\tilde{\mathbf{\Sigma}}^{*}_{\mathrm{A},\hat{k}}$.
\vspace{-3mm}
\subsection{Proof of Proposition \ref{prop:bound}}\label{app:B}
Since problem \eqref{eq:sca} is Lipschitz continuous and a quadratic programs problem, solving problem \eqref{eq:sca} is to find the point in the feasible region that is closest to the stationary point, as
\begin{equation}\label{eq:opt}
	\mathbf{r}_{m} = \mathop{\arg\min}\limits_{\mathbf{r}_{m} \in \Omega_{n}} \frac{1}{2} \Vert \mathbf{r}_{m} - \mathbf{r}^{\star}_{m} \Vert^{2}_{2}.
\end{equation}
We prove Proposition \ref{prop:bound} by performing a thorough analysis on the KKT conditions of problem \eqref{eq:opt}. The KKT condition of problem \eqref{eq:opt} as follows:
\begin{subequations}
	\begin{align}
		& \left(\mathbf{r}_{m} - \mathbf{r}^{\star}_{m}\right) - \sum_{i=1}^{N_{\mathrm{A}}-1}\varpi_{i}\nabla g_{i} + \sum_{j=1}^{4}\varsigma_{j}\nabla h_{j} = 0 \label{eq:foo} \\
		& g_{i}\left(\mathbf{r}_{m}\right) \geq 0, \ \forall i, \ h_{j}\left(\mathbf{r}_{n}\right) \leq 0, \ \forall j \label{eq:csc}\\
		& \varpi_{i}g_{i} = 0, \ \forall i, \ \varsigma_{j}h_{j} = 0, \ \forall j \label{eq:primal}\\
		&  \varpi_{i} \geq 0, \ \forall i, \ \varsigma_{j} \geq 0, \ \forall j \label{eq:dual}
	\end{align}
\end{subequations}
where $g_{i} = \Vert \mathbf{r}_{m} - \mathbf{r}_{i} \Vert^{2}_{2} - D^{2}$ and $h_{1} = x^{\mathrm{r}}_{m} - A_{\mathrm{r}}/2, h_{2} = -x^{\mathrm{r}}_{m} - A_{\mathrm{r}}/2, h_{3} = y^{\mathrm{r}}_{m} - A_{\mathrm{r}}/2, h_{4} = -y^{\mathrm{r}}_{m} - A_{\mathrm{r}}/2$. \eqref{eq:foo} is the first-order optimality conditions; \eqref{eq:csc} is the complementary slackness condition; \eqref{eq:primal} and \eqref{eq:dual} are the primal and dual feasibility conditions, respectively. If $\mathbf{r}_{m}$ in interior of $\Omega_{m}$, we can get that $ g_{i}\left(\mathbf{r}_{m}\right) > 0$ and $h_{j}\left(\mathbf{r}_{m}\right) < 0$, which means $\mathbf{r}_{m} = \mathbf{r}^{\star}_{m}$. It is contradictory to $ \mathbf{r}^{\star}_{m} \notin \Omega_{n}$. So $\mathbf{r}_{m}$ must satisfy at least one of the boundary conditions of the constraints.
\par
Define \(\mathbf{r}_{\mathrm{proj}}\) as the projection of \(\mathbf{r}^{\star}_{m}\) onto the square \(\{|x| \leq A_{\mathrm{r}}/2, |y| \leq A_{\mathrm{r}}/2\}\). In the following cases, we will prove that if \(\mathbf{r}_{\mathrm{proj}}\) does not satisfy the minimum distance constraint, then $\mathbf{r}_{m}$ must lie on at least one of the minimum distance boundaries.
\par
\subsubsection*{Case 1: \(\mathbf{r}^{\star}_{n}\) Outside Square, \(\mathbf{r}_{\mathrm{proj}}\) Does not Violate Distance Constraint} 
\par
If \(\|\mathbf{r}_{\mathrm{proj}} - \mathbf{r}_{i}\|^{2}_{2} \geq D^2, \forall i\), then \(\mathbf{r}_{\mathrm{proj}} \in \Omega_n\). Since the objective is convex and \(\mathbf{r}_{\mathrm{proj}}\) is the closest point in the square to \(\mathbf{r}^{\star}_{m}\), \(\mathbf{r}_{m} = \mathbf{r}_{\mathrm{proj}}\).

\subsubsection*{Case 2: \(\mathbf{r}^{\star}_{m}\) Outside Square, \(\mathbf{r}_{\mathrm{proj}}\) Violates Distance Constraint} 
\par
Assume by contradiction that \(g_i(\mathbf{r}_{m}) > 0, \forall i\), so \(\varpi_i = 0, \forall i\), and the KKT condition reduces to:
\begin{equation}
	\mathbf{r}_{m} - \mathbf{r}^{\star}_{m} + \sum_{j=1}^{4} \varsigma_j \nabla h_j = 0.
\end{equation}

If \(\exists i\) such that \(\|\mathbf{r}_{\mathrm{proj}} - \mathbf{r}_{i}\|^{2}_{2} < D^2\), then \(\mathbf{r}_{\mathrm{proj}} \notin \Omega_m\). Consider:
\begin{itemize}
	\item \textit{Subcase 1: \(\mathbf{r}_{m}\) Inside Square}
	
	 If \(\varsigma_j = 0, \forall j\), then \(\mathbf{r}_{m} = \mathbf{r}^{\star}_{m}\), contradicting \(\mathbf{r}^{\star}_{m}\) being outside.
	\item \textit{Subcase 2: \(\mathbf{r}_{m}\) on Square Boundary}
	
	Let \(\mathbf{r}^{\star}_{m} = (x^{\star}, y^{\star})\) with \(x^{\star} > A_{\mathrm{r}}/2\). If \(\mathbf{r}_{m} = (A_{\mathrm{r}}/2, y^{\mathrm{r}}_{m})\) (\(h_1 = 0\)), then:
	\begin{equation}
		(A_{\mathrm{r}}/2 - x^{\star}, y^{\mathrm{r}}_{m} - y^{\star}) + \varsigma_1 (1, 0) = 0
	\end{equation}
	so \(\varsigma_1 = x^{\star} - A_{\mathrm{r}}/2 > 0\), \(y^{\mathrm{r}}_{m} = y^{\star}\). Thus, \(\mathbf{r}_{m} = (A_{\mathrm{r}}/2, y^{\star}) = \mathbf{r}_{\mathrm{proj}}\), but \(\mathbf{r}_{\mathrm{proj}} \notin \Omega_m\), contradicting \(g_i(\mathbf{r}_{m}) > 0\).
	\item \textit{Other Boundary Subcases:}  
	The remaining cases (e.g., \(\mathbf{r}_m\) on other edges/corners of the square) follow analogously by symmetry.
	 We omit repetitive details for brevity.
\end{itemize}
Hence, \(g_i > 0, \forall i\) fails, and \(\exists i, g_i(\mathbf{r}_{m}) = 0\).
\par
\subsubsection*{Case 3: \(\mathbf{r}^{\star}_{m}\) Inside Square, Violates Distance Constraint}
If \(|x^{\star}| \leq A_{\mathrm{r}}/2\), \(|y^{\star}| \leq A_{\mathrm{r}}/2\), but \(\exists i, \|\mathbf{r}^{\star}_{m} - \mathbf{r}_{i}\|^{2}_{2} < D^2\), then \(\mathbf{r}^{\star}_{m} \notin \Omega_m\). Consider:
\begin{itemize}
	\item \textit{Subcase 1: \(\mathbf{r}_{m}\) Inside Square}: If \(\varsigma_j = 0, \forall j\), then \(\mathbf{r}_{m} = \mathbf{r}^{\star}_{m}\), contradicting \(\mathbf{r}^{\star}_{m} \notin \Omega_m\).
	\item \textit{Subcase 2: \(\mathbf{r}_{m}\) on Square Boundary}: Let \(\mathbf{r}_{m} = (A_{\mathrm{r}}/2, y_{m})\) (\(h_1 = 0\)):
	\begin{equation}
		(A_{\mathrm{r}}/2 - x^{\star}, y_{m} - y^{\star}) + \varsigma_1 (1, 0) = 0
	\end{equation}
	so \(\varsigma_1 = x^{\star} - A_{\mathrm{r}}/2\). Since \(x^{\star} \leq A_{\mathrm{r}}/2\), \(\varsigma_1 \leq 0\). If \(x^{\star} < A_{\mathrm{r}}/2\), \(\varsigma_1 < 0\), contradicting \(\varsigma_1 \geq 0\). If \(x^{\star} = A_{\mathrm{r}}/2\), \(\varsigma_1 = 0\), then \(y_{m} = y^{\star}\), so \(\mathbf{r}_{m} = (A_{\mathrm{r}}/2, y^{\star}) = \mathbf{r}^{\star}_{m}\), contradicting \(\mathbf{r}^{\star}_{m} \notin \Omega_m\).
	\item \textit{Other Boundary Subcases:}  
	The remaining cases (e.g., \(\mathbf{r}_m\) on other edges/corners of the square) follow analogously by symmetry.
	We omit repetitive details for brevity.
	
\end{itemize}
Thus, \(g_i > 0, \forall i\) leads to a contradiction, and \(\exists i, g_i(\mathbf{r}_{m}) = 0\).
\par
In $\textit{Case} \ 2$ and $\textit{Case} \ 3$, \(\mathbf{r}_{m}\) lies on some \(g_i = 0\) boundary.

\vspace{-3mm}

\bibliographystyle{IEEEtran} 


\end{document}